\documentclass[journal]{IEEEtran}

\usepackage{cite}
\usepackage{amsmath}
\usepackage{verbatim}
\usepackage{graphics}
\usepackage{algorithm}
\usepackage{algorithmic}
\usepackage[pdftex]{graphicx}
\usepackage{epstopdf}
\usepackage{subfigure}
\usepackage{float}
\usepackage{multirow}
\usepackage{stfloats}
\usepackage{array}
\usepackage{times}
\usepackage{epsfig}
\usepackage{graphicx}
\usepackage{color}
\usepackage{booktabs}
\usepackage{comment}
\usepackage{threeparttable}
\usepackage{amssymb}
\usepackage{url}
\makeatletter
\let\NAT@parse\undefined
\makeatother
\usepackage{hyperref}

\begin{document}

\title{Real-World Single Image Super-Resolution: \\  A Brief Review }

\author{Honggang Chen,
    	Xiaohai He,
    	Linbo Qing, 
    	Yuanyuan Wu,
     	Chao Ren, 
    	and Ce Zhu		
	\thanks{This work was supported by the National Natural Science Foundation of China under Grant 62001316, Grant 61801316, and Grant 61871279. (\emph{Corresponding author: Chao Ren.})}
	\thanks{Honggang Chen, Xiaohai He, Linbo Qing, and Chao Ren are with the College of Electronics and Information Engineering, Sichuan University, Chengdu 610065, China (e-mail: honggang\_chen@scu.edu.cn; hxh@scu.edu.cn; qing\_lb@scu.edu.cn; chaoren@scu.edu.cn). }
	\thanks{Yuanyuan Wu is with the College of Information Science and Technology, Chengdu University of Technology, Chengdu, China, 610059 (e-mail: wuyuanyuan@cdut.edu.cn). }
    \thanks{Ce Zhu is with the School of Information and Communication Engineering, University of Electronic Science and Technology of China, Chengdu 611731, China (e-mail: eczhu@uestc.edu.cn)}}

\markboth{Manuscript Submitted to XXX}
{Chen \MakeLowercase{\textit{et al.}}: XXXX}
\maketitle

\begin{abstract}
Single image super-resolution (SISR), which aims to reconstruct a high-resolution (HR) image from a low-resolution (LR) observation, has been an active research topic in the area of image processing in recent decades. Particularly, deep learning-based super-resolution (SR) approaches have drawn much attention and have greatly improved the reconstruction performance on synthetic data. Recent studies show that simulation results on synthetic data usually overestimate the capacity to super-resolve real-world images. In this context, more and more researchers devote themselves to develop SR approaches for realistic images. This article aims to make a comprehensive review on real-world single image super-resolution (RSISR). More specifically, this review covers the critical publically available datasets and assessment metrics for RSISR, and four major categories of RSISR methods, namely the degradation modeling-based RSISR, image pairs-based RSISR, domain translation-based RSISR, and self-learning-based RSISR. Comparisons are also made among representative RSISR methods on benchmark datasets, in terms of both reconstruction quality and computational efficiency. Besides, we discuss challenges and promising research topics on RSISR. 
\end{abstract}


\IEEEpeerreviewmaketitle
\section{Introduction}
\label{S.Introd}
\IEEEPARstart{H}{igh-resolution (HR)}  images are desired urgently in many application areas such as intelligent surveillance, medical imaging, and remote sensing. To obtain images with higher resolution, a natural idea is to upgrade the hardware (\emph{e.g.}, the imaging system). Although recent years have witnessed the obvious progress of imaging devices and techniques, this kind of approach has two main limitations: (\romannumeral1) It is inflexible and costly because the demand in practical applications is constantly changing. (\romannumeral2) It can be used only for capturing new HR images, but not for enhancing the resolution of existing low-resolution (LR) images. Compared to the hardware upgrade-based ``hard" solution, the signal processing-based ``soft" image resolution enhancement known as super-resolution (SR) is more flexible and economical. With the SR techniques that reconstruct a higher resolution output from the LR observation, we can obtain images with the resolution beyond the limit of imaging systems, thereby benefiting the subsequent analysis and understanding tasks such as segmentation \cite{dai2016image, lei2019simultaneous, guo2019super, wang2020dual}, detection \cite{haris2018task, pang2019jcs, zhang2020kgsnet}, and recognition \cite{wang2016studying, yang2018long, suprapto2019gsr}.
\begin{figure}[!tb]
	\centering
	\includegraphics[width = 9 cm]{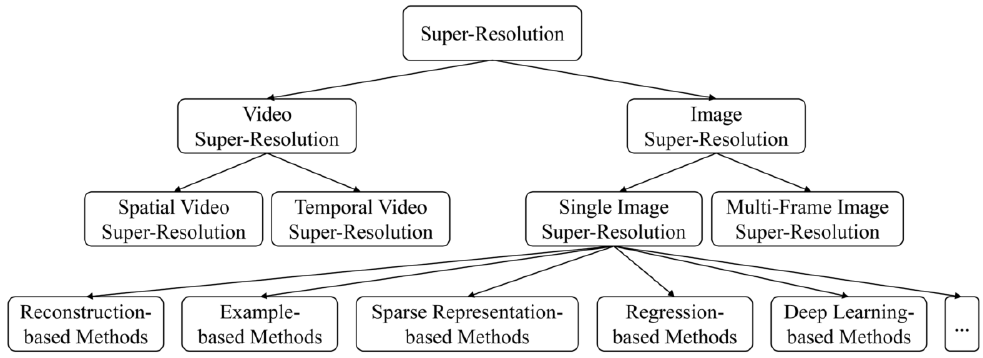}
	\caption{The taxonomy of existing super-resolution techniques.}
	\label{Fig.1}
\end{figure}

In general, as presented in Fig. \ref{Fig.1}, existing SR techniques can be grouped into two categories according to the LR input and the reconstructed HR output, \emph{i.e.}, video super-resolution (VSR) \cite{belekos2010maximum, liu2013bayesian, li2016video, kappeler2016video, caballero2017real, jo2018deep, lucas2019generative, haris2019recurrent, tian2020tdan, dikbas2012novel, choi2007motion, niklaus2017video1, niklaus2018context, bao2019depth, peleg2019net, bao2019memc, cheng2020video} and image super-resolution (ISR) \cite{farsiu2004fast, farsiu2005multiframe, li2010multi, yuan2011multiframe, yue2014locally, kohler2016robust, liu2018robust, laghrib2019new, sun2008image, zhang2012single, dong2012nonlocally, papyan2015multi, jiang2016single, ren2016single, chen2017single, chang2019data, li2019single, li2020adaptive,  freeman2002example, glasner2009super, xiong2013example, zhu2014single, huang2015single, li2016rotation, huang2017srhrf+, yang2010image, zeyde2010single, wang2012semi, zhu2014fast, kang2015learning, li2020combining, li2020depth, ayas2020single, timofte2013anchored, timofte2014a+, zhang2015joint, agustsson2016regressor, perez2016antipodally, zhang2019learning, dong2015image, kim2016accurate, ledig2017photo, lim2017enhanced, haris2018deep, lai2018fast, zhang2018image, zhang2018residual, dai2019second, guo2020closed, 2020Residual, zhang2020gated}. On the whole, VSR aims to improve the spatial resolution (known as spatial VSR) \cite{belekos2010maximum, liu2013bayesian, li2016video, kappeler2016video, caballero2017real, jo2018deep, lucas2019generative, haris2019recurrent, tian2020tdan} or the frame rate (known as temporal VSR) \cite{dikbas2012novel, choi2007motion, niklaus2017video1, niklaus2018context, bao2019depth, peleg2019net, bao2019memc, cheng2020video} of the observed video. ISR can be further classified into multi-frame image super-resolution (MISR) \cite{farsiu2004fast, farsiu2005multiframe, li2010multi, yuan2011multiframe, yue2014locally, kohler2016robust, liu2018robust, laghrib2019new} and single image super-resolution (SISR) \cite{sun2008image, zhang2012single, dong2012nonlocally, papyan2015multi, jiang2016single, ren2016single, chen2017single, chang2019data, li2019single, li2020adaptive, freeman2002example, glasner2009super, xiong2013example, zhu2014single, huang2015single, li2016rotation, huang2017srhrf+, yang2010image, zeyde2010single, wang2012semi, zhu2014fast, kang2015learning, li2020combining, li2020depth, ayas2020single, timofte2013anchored, timofte2014a+, zhang2015joint, agustsson2016regressor, perez2016antipodally, zhang2019learning, dong2015image, kim2016accurate, ledig2017photo, lim2017enhanced, haris2018deep, lai2018fast, zhang2018image, zhang2018residual, dai2019second, guo2020closed, 2020Residual, zhang2020gated}.  MISR refers to reconstructing an HR image via fusing the complementary information in a series of correlated images of the same scene  \cite{farsiu2004fast, farsiu2005multiframe, li2010multi, yuan2011multiframe, yue2014locally, kohler2016robust, liu2018robust, laghrib2019new}, while SISR generates an HR image from one LR observation \cite{sun2008image, zhang2012single, dong2012nonlocally, papyan2015multi, jiang2016single, ren2016single, chen2017single, chang2019data, li2019single, li2020adaptive, freeman2002example, glasner2009super, xiong2013example, zhu2014single, huang2015single, li2016rotation, huang2017srhrf+, yang2010image, zeyde2010single, wang2012semi, zhu2014fast, kang2015learning, li2020combining, li2020depth, ayas2020single, timofte2013anchored, timofte2014a+, zhang2015joint, agustsson2016regressor, perez2016antipodally, zhang2019learning, dong2015image, kim2016accurate, ledig2017photo, lim2017enhanced, haris2018deep, lai2018fast, zhang2018image, zhang2018residual, dai2019second, guo2020closed, 2020Residual, zhang2020gated}. In terms of application scenarios, SISR is more practical than MISR and VSR because it is much less demanding on the input, which is one reason why SISR attracts wider attention. A variety of SISR methods have been proposed in the past decades, mainly including reconstruction-based \cite{sun2008image, zhang2012single, dong2012nonlocally, papyan2015multi, jiang2016single, ren2016single, chen2017single, chang2019data, li2019single, li2020adaptive}, example-based \cite{freeman2002example, glasner2009super, xiong2013example, zhu2014single, huang2015single, li2016rotation, huang2017srhrf+}, sparse representation-based \cite{yang2010image, zeyde2010single, wang2012semi, zhu2014fast, kang2015learning, li2020combining, li2020depth, ayas2020single}, regression-based \cite{timofte2013anchored, timofte2014a+, zhang2015joint, agustsson2016regressor, perez2016antipodally, zhang2019learning}, and deep learning-based approaches \cite{dong2015image, kim2016accurate, ledig2017photo, lim2017enhanced, haris2018deep, lai2018fast, zhang2018image, zhang2018residual, dai2019second, guo2020closed, 2020Residual, zhang2020gated}, \emph{etc}. Particularly, the deep learning-based SISR methods \cite{dong2015image, kim2016accurate, ledig2017photo, lim2017enhanced, haris2018deep, lai2018fast, zhang2018image, zhang2018residual, dai2019second, guo2020closed, 2020Residual, zhang2020gated} developed in recent years take the SR performance on synthetic LR images (\emph{e.g.}, bicubically downsampled images) to a new level. 

Nevertheless, previous research \cite{kohler2019toward} shows that the actual SR ability of most existing SISR methods may be overestimated based only on synthetic data due to the domain gap between synthetic and realistic data. In other words, compared to the promising SR results on synthetic test images,  the SR performance would degrade significantly on real-world images, thus hindering the practical applications of SISR algorithms. To address this problem, some researchers have shifted their focus to real-world single image SR (RSISR) over the past couple of years, and a series of studies involving real-world dataset collection \cite{cai2019toward, wei2020component, chen2019camera, zhang2019zoom, wang2020scene, kohler2019toward, reza2020imagepairs,xu2019towards, xu2020eploiting}, SR models for real-world images \cite{shao2015simple, shao2019nonparametric, gu2019blind, cornillere2019blind, huang2020unfolding, michaeli2013nonparametric, bell2019blind, bulat2018learn, zhou2019kernel, xiao2020degradation, ji2020real, cai2019toward, zhang2019zoom, xu2019towards, xu2020eploiting,  chen2019camera, wang2020scene, kohler2019toward, wei2020component, reza2020imagepairs, yuan2018unsupervised, zhang2019multiple, kim2020unsupervised, maeda2020unpaired, prajapati2020unsupervised, zhao2018unsupervised, you2019ct, fritsche2019frequency, muhammad2020deep, rad2021benefiting, lugmayr2019unsupervised, chen2020unsupervised, shocher2018zero, kim2020dual, emad2021dualsr, soh2020meta, park2020fast}, and SR result assessment \cite{ma2017learning, fang2018blind, bare2018deep, greeshma2020super} have been conducted. Meanwhile, several challenges on RSISR have been organized in conjunction with the \emph{IEEE Conference on Computer Vision and Pattern Recognition (CVPR)}, \emph{IEEE International Conference on Computer Vision (ICCV)}, and \emph{European Conference on Computer Vision (ECCV)} to attract more attention and promote the development of RSISR techniques \cite{cai2019ntire, lugmayr2019aim, lugmayr2020ntire, wei2020aim}. It is exciting to see that the studies on RSISR are becoming more targeted and the SR performance on real-world images is improved increasingly. 

In this work, we mainly give an overview of recent RSISR algorithms and relevant studies. There are several works concerning the overview of image and video SR techniques. For example,  Yue \emph{et al.} \cite{yue2016image} make a summary of the techniques, applications and future of image SR. Considering that deep learning has been widely employed to address SR over the past few years, more recently Yang \emph{et al.} \cite{yang2019deep}, Wang \emph{et al.} \cite{wang2020deep}, and Liu \emph{et al.} \cite{liu2020video} review deep learning-based image/video SR methods. However, this work is the first attempt to make an overview of RSISR techniques to the best of our knowledge. The main contributions of this review are four-fold: (\romannumeral1) We comprehensively review the studies on RSISR, including datasets, assessment metrics, technologies and methods, \emph{etc}. (\romannumeral2) We present a taxonomy for existing RSISR methods according to their primary principles. (\romannumeral3) We compare the reconstruction accuracy and efficiency of representative RSISR algorithms on benchmark datasets. (\romannumeral4) We further discuss current challenges and future research directions for RSISR.

The rest of this review is organized as follows. The background of RSISR is briefly introduced in Section \uppercase\expandafter{\romannumeral2}. In Section \uppercase\expandafter{\romannumeral3}, the datasets and assessment metrics for RSISR are described. Section \uppercase\expandafter{\romannumeral4} reviews RSISR technologies and methods by category. The comparisons among representative RSISR algorithms are presented in Section \uppercase\expandafter{\romannumeral5}. In Section \uppercase\expandafter{\romannumeral6}, we analyze current challenges and future research directions of RSISR. Finally, Section \uppercase\expandafter{\romannumeral7} concludes this work.

\section{Background}
\label{S.Backg}
SISR refers to reconstructing an HR image from an LR observation. Given an LR image ${\bf{Y}}$, it is generally assumed to be degraded from a corresponding HR image ${\bf{X}}$, which can be represented as
\begin{equation}\label{eq.1}
{\bf{Y}} = D({\bf{X}},{\theta _D})
\end{equation} 
where $D(\cdot)$ denotes the degradation process defined by the parameter set ${\theta _D}$. Note that in a real scenario, the degradation parameter ${\theta _D}$ is unknown, and all we have is the LR image ${\bf{Y}}$. SISR aims at
recovering a good estimate of the potential HR image via reversing the degradation process shown in Eq.~(\ref{eq.1}), which can be formulated as 
\begin{equation}\label{eq.2}
{\bf{\hat X}} = R({\bf{Y}},{\theta _R})
\end{equation} 
where $R(\cdot)$ represents the SR function and ${\theta _R}$ is the corresponding parameter set. ${\bf{\hat X}}$ is the super-resolved image from ${\bf{Y}}$, \emph{i.e.}, an estimate of the real HR image ${\bf{X}}$.

Apparently, the SR process and degradation process are the inverses of each other. Thus, for obtaining excellent reconstruction performance, the SR function $R({\bf{Y}},{\theta _R})$ should be adapted to the degradation  $D({\bf{X}},{\theta _D})$. In the literature, some researchers \cite{sun2008image, zhang2012single, dong2012nonlocally, papyan2015multi, jiang2016single, ren2016single, chen2017single, chang2019data, li2019single, li2020adaptive} approximate the degradation via blurring, downsampling, and noise injection. Mathematically, the simulated degradation process is as follows
\begin{equation}\label{eq.3}
{\bf{Y}} = {\bf{SBX}} + {\bf{n}}
\end{equation} 
where $\bf{B}$ and $\bf{S}$ denote the operations of blurring and downsampling, respectively. In general, the blurring is realized via convolving the HR image with a Gaussian kernel. $\bf{n}$ represents the additive noise, which is usually assumed to be white Gaussian noise. Part of works \cite{timofte2013anchored, timofte2014a+, zhang2015joint, agustsson2016regressor, perez2016antipodally, zhang2019learning, dong2015image, kim2016accurate, ledig2017photo, lim2017enhanced, haris2018deep, lai2018fast, zhang2018image, zhang2018residual, dai2019second, guo2020closed, 2020Residual, zhang2020gated} adopt a simpler degradation model, \emph{i.e.}, directly downscaling an HR image using the ``bicubic'' kernel to generate corresponding LR image. For comparison and evaluation, most existing SR methods are developed and validated on synthetic LR images generated by degradation simulations. Overall, the SR reconstruction performance on synthetic LR images is rather good, especially for deep learning-based SISR approaches such as RCAN \cite{zhang2018image}, SAN \cite{dai2019second}, and RFANet \cite{2020Residual}.

Compared with the commonly used degradation model in simulations, the actual degradation in real-world scenarios is more complex and varying because it can be affected by a number of factors (\emph{e.g.}, imaging system and imaging environment). In other words, the degradation model used in simulations may not match that suffered by real-world images, which results in the domain gap between synthetic LR images and realistic LR observations. For this primary reason, the reconstruction performance of most existing SISR algorithms drops significantly on real-world images. To enhance quality of super-resolving real-world images, some researchers have been working on RSISR from different perspectives in the past several years, including realistic dataset building, SR model development, SR performance assessment,~\emph{etc}. 

\section{Datasets and Assessment Metrics}
\label{S.DataEvalMe}
Training/testing datasets and assessment metrics are the cornerstones of the SISR. In this section, we briefly introduce the relevant datasets and assessment metrics.

\subsection{Datasets for RSISR}
\label{S.DataSet}
For the training and testing of SISR models, the widely used datasets include  DIV2K \cite{agustsson2017ntire}, BSDS500 \cite{arbelaez2011contour},  T91 \cite{yang2010image}, Set5 \cite{bevilacqua2012low}, Set14 \cite{zeyde2010single}, Urban100 \cite{huang2015single}, Manga109 \cite{fujimoto2016manga109}, \emph{etc}. Most of these datasets only contain HR images. In this case, we need to generate LR counterparts based on the assumed degradation model (\emph{e.g.}, ``bicubic'' kernel-based downsampling), both for model training and testing. Therefore, these datasets are not quite suitable for the study of RSISR due to the significant discrepancy between the assumed degradation model and the real one. To address this problem, some more targeted datasets for RSISR have been constructed in different ways, including DIV2KRK \cite{bell2019blind}, RealSR  \cite{cai2019toward}, DRealSR \cite{wei2020component},   City100 \cite{chen2019camera}, SR-RAW \cite{zhang2019zoom}, TextZoom \cite{wang2020scene}, SupER \cite{kohler2019toward},  ImagePairs \cite{reza2020imagepairs},~\emph{etc}. Table \ref{Tab.1} summarizes the above datasets. 

\begin{table}[!t]
	\center
	\caption{An overview of datasets for RSISR.}
	\scalebox{0.825}[0.825]{
	\begin{tabular}{m{1.1cm}<{\centering}m{1.7cm}<{\centering}m{1.2cm}<{\centering}m{1cm}<{\centering}m{3.5cm}<{\centering}}
		\hline
		Datasets          & Published                   & Synthetic / Realistic     & Scale Factors  & Keywords \\ \hline  	\hline
		DIV2KRK           & NeurIPS-2019 \cite{bell2019blind}       &  Synthetic    & $\times$2, $\times$4      &   DIV2K, Random kernels, Uniform multiplicative noise       \\ \hline
		RealSR          & ICCV-2019 \cite{cai2019toward}            &  Realistic    &  $\times$2, $\times$3, $\times$4     &   Focal length adjusting   \\ \hline
		DRealSR          & ECCV-2020 \cite{wei2020component}        &  Realistic    &  $\times$2, $\times$3, $\times$4     &   Focal length adjusting        \\ \hline
		City100          & CVPR-2019 \cite{chen2019camera}          &  Realistic    &  $\times$2.9, $\times$2.4      &    Focal length adjusting, Shooting distance changing       \\ \hline
		SR-RAW          & CVPR-2019 \cite{zhang2019zoom}            &  Realistic    & $\times$4, $\times$8      &   Focal length adjusting, RAW data       \\ \hline
		TextZoom          &  ECCV-2020 \cite{wang2020scene}         &  Realistic    &   $\times$2    &  Text, Recognition        \\ \hline
		SupER          & TPAMI-2020 \cite{kohler2019toward}         &  Realistic    &  $\times$2, $\times$3, $\times$4     &   Hardware binning, Image sequences        \\ \hline
		ImagePairs          &  CVPRW-2020 \cite{reza2020imagepairs} &  Realistic    &  $\times$2     &     Beam-splitter cube, RAW data     \\ \hline
	\end{tabular}}
	\label{Tab.1}
\end{table}

\subsubsection{DIV2KRK \cite{bell2019blind}}  This is a synthetic testing dataset built by Bell-Kligler \emph{et al.} \cite{bell2019blind} for blind SR. \footnote{DIV2KRK \cite{bell2019blind} is available at \url{http://www.wisdom.weizmann.ac.il/~vision/kernelgan/}}  The DIV2KRK \cite{bell2019blind} is derived from DIV2K \cite{agustsson2017ntire} that contains diverse 2K resolution images. DIV2K is proposed by Agustsson \emph{et al.} \cite{agustsson2017ntire} for NTIRE 2017 SR challenge. More specifically, the 100 HR images in the validation set of DIV2K \cite{agustsson2017ntire} are blurred and downsampled with random kernels to generate LR counterparts. Random blur kernels are $11 \times 11$ anisotropic gaussians, and each of them is with two independently distributed lengths ${\lambda_1},{\lambda_2}\sim u(0.6,5)$ and a rotation angle $\theta \sim u\left[{-\pi,\pi}\right]$. Further, uniform multiplicative noise is applied to the blur kernel before normalizing its sum to one. DIV2KRK \cite{bell2019blind} is still a synthetic dataset, although the degradation model is  more complex and random.

\subsubsection{RealSR \cite{cai2019toward}} RealSR is a real-world dataset collected by Cai~\emph{et al.} \cite{cai2019toward} for training and testing  RSISR models. \footnote{RealSR \cite{cai2019toward} is available at \url{https://github.com/csjcai/RealSR}}
RealSR consists of 595 LR-HR image pairs obtained by adjusting the lens of two digital single lens reflex (DSLR) cameras (\emph{i.e.}, Nikon D810 and Canon 5D3). To study the RSISR with different scaling factors (\emph{e.g.}, $\times$2, $\times$3, $\times$4), images  are taken at four focal lengths, \emph{i.e.}, 28mm, 35mm, 50mm, and 105mm.  Naturally, for the same scene, the images captured at 105mm are used as HR images, and the images captured at 28mm, 35mm, and 50mm are seen as LR counterparts for $\times$4, $\times$3, and $\times$2 upsampling, respectively. Considering the differences (\emph{e.g.}, lens distortion and exposures) among the images captured at different focal lengths, Cai \emph{et al.} \cite{cai2019toward} further propose a progressive image registration framework to achieve pixel-wise registration of images taken at 28mm, 35mm, 50mm, and 105mm. First, PhotoShop is used to correct the lens distortion and the regions of interest around the center of the corrected images are cropped. Then, the regions cropped from the images captured at 105mm are used as HR references, and the corresponding regions cropped from the images taken at 28mm, 35mm, and 50mm are aligned to generate LR counterparts via iteratively optimizing the parameters of affine transformation and luminance adjustment. After convergence, real-world LR-HR image pairs can be obtained. 

\subsubsection{DRealSR \cite{wei2020component}} The real-world dataset DRealSR built by Wei~\emph{et al.} \cite{wei2020component}  is similar to RealSR \cite{cai2019toward}, with a larger scale. \footnote{DRealSR \cite{wei2020component} is available at \url{https://github.com/xiezw5/Component-Divide-and-Conquer-for-Real-World-Image-Super-Resolution}}
More specifically, five DLSR cameras (\emph{i.e.}, Sony, Canon, Olympus, Nikon, and Panasonic) are used to capture images at four resolutions in outdoor and indoor scenes (\emph{e.g.}, buildings, offices, plants, posters, \emph{etc.}). The SIFT algorithm \cite{lowe2004distinctive} is adopted to align the images with different resolutions. In total, DRealSR \cite{wei2020component} contains 884, 783, and 840 LR-HR image pairs for $\times$2, $\times$3, and $\times$4 SR, respectively.

\subsubsection{City100 \cite{chen2019camera}} The City100 dataset proposed by Chen \emph{et al.} \cite{chen2019camera} includes City100\_NikonD5500 and City100\_iPhoneX, which characterize the resolution and field-of-view (FoV) degradation under DSLR and smartphone cameras, respectively. \footnote{City100 \cite{chen2019camera} is available at \url{https://github.com/ngchc/CameraSR}} As is well known, there is an tradeoff between the resolution and FoV for imaging systems. An image with a larger FoV but a lower resolution can be obtained when zooming out the lens, while the image resolution can be enhanced at the expense of a reduced FoV when zooming in the lens. Therefore, Chen \emph{et al.} propose \cite{chen2019camera} to adjust the focal length or shooting distance to capture images of the same scene with different resolutions. 100 postcards of different city scenes are used as imaging subjects. For the City100\_NikonD5500 dataset taken by NikonD5500, HR and LR images are captured at the focal length of 55mm and 18mm, respectively. The HR and LR images of the same scene are first spatially aligned based on SIFT key-points \cite{lowe2004distinctive} and RANSAC  \cite{fischler1981random}. Further, intensity and color rectification is conducted to improve the accuracy of alignment. The main distinction between City100\_iPhoneX and City100\_NikonD5500 is that City100\_iPhoneX is captured by iPhone X via changing the shooting distance.

\subsubsection{SR-RAW \cite{zhang2019zoom}}  The SR-RAW dataset proposed by Zhang~\emph{et al.} \cite{zhang2019zoom} is composed of pairs of RAW images taken at different levels of optical zoom. \footnote{SR-RAW \cite{zhang2019zoom} is available at \url{https://ceciliavision.github.io/project-pages/project-zoom.html}} The SR-RAW \cite{zhang2019zoom} is similar to the RealSR \cite{cai2019toward} mentioned above in terms of the way to capture images of the same scene with different resolutions, \emph{i.e.}, adjusting the focal length. For SR-RAW \cite{zhang2019zoom}, seven images of each scene are taken under seven different optical zoom settings using a 24-240mm zoom lens (\emph{i.e.}, Sony FE 24-240mm). In total, 500 seven-image sequences are collected in outdoor and indoor scenes. For image registration, the Euclidean motion model is applied to describe the relationship between the images with different resolutions and it is optimized by minimizing the enhanced correlation coefficient as in  \cite{evangelidis2008parametric}. Unlike the RealSR built by Cai \emph{et al.} \cite{cai2019toward}, SR-RAW \cite{zhang2019zoom} contains both raw sensor data and RGB images because it is used for SR from raw data. That is, instead of the LR RGB image, its raw sensor data is used as the input to reconstruct the corresponding HR RGB image. Actually, the reconstruction process covers demosaicing, denoising, SR, \emph{etc}. Compared with 8-bit RGB images, as pointed out in \cite{zhang2019zoom}, raw sensor data generally contains more useful information for SR.

\subsubsection{TextZoom \cite{wang2020scene}}  The TextZoom derived from RealSR \cite{cai2019toward} and SR-RAW \cite{zhang2019zoom} is the first real scene text SR dataset constructed by Wang \emph{et al.}  \cite{wang2020scene}. \footnote{TextZoom \cite{wang2020scene} is available at \url{https://github.com/JasonBoy1/TextZoom}}. More specifically, the text images in TextZoom  \cite{wang2020scene} are cropped from the images in RealSR \cite{cai2019toward} and SR-RAW \cite{zhang2019zoom}, including various natural scenes such as shops, street views, and vehicle interiors. The content, direction, and focal length of each LR-HR text image pair in TextZoom are provided in the annotation process. Moreover,  TextZoom \cite{wang2020scene} comprises three subsets according to difficulty levels, namely easy, medium, and hard. Thanks to the well-organized annotations, TextZoom \cite{wang2020scene} can be utilized to study text image SR as well as text recognition.

\subsubsection{SupER \cite{kohler2019toward}} The SupER is built by K{\"o}hler \emph{et al.} \cite{kohler2019toward} via hardware binning. \footnote{SupER \cite{kohler2019toward} is available at \url{https://www.lms.tf.fau.eu/research/downloads/superresolution/}}
More specifically, more than 80,000 images are taken from 14 lab scenes at four imaging resolutions and five compression levels, using a Basler acA2000-50gm CMOS camera with a $f$/1.8, 16mm fixed-focus lens.  What is special about SupER \cite{kohler2019toward} is that the imaging resolution is adjusted by changing the binning factor, which naturally guarantees the perfect alignment between LR and HR images. Three binning factors (\emph{i.e.}, 2, 3, and 4) are used to acquire LR images corresponding to the same HR image at three different resolution levels. Meanwhile, to enhance the comprehensiveness, four motion types, two photometric conditions, and five H.265 compression levels are considered while building SupER \cite{kohler2019toward}. Additionally, unlike most existing datasets for SISR, image sequences instead of independent images are captured in SupER \cite{kohler2019toward}, which is also applicable in the study on MISR algorithms.

\subsubsection{ImagePairs \cite{reza2020imagepairs}} The ImagePairs proposed by Joze \emph{et al.} \cite{reza2020imagepairs} includes 11,421 LR-HR image pairs of diverse scenes, in which LR and HR images are captured by an LR camera (5MP) and an HR camera (20.1MP), respectively. \footnote{ImagePairs \cite{reza2020imagepairs} is available at \url{www.microsoft.com/en-us/research/project/imagepairs}}  More specifically, a beam-splitter cube is used to make the two cameras capture images of the same scene simultaneously. Due to the difference in focal length, the LR and HR cameras have different perspectives. Therefore, Joze \emph{et al.} \cite{reza2020imagepairs} propose to generate pixel-wise aligned LR-HR image pairs via applying the following four steps: ISP, image undistortion, pair alignment, and margin cropping. The raw data collected by LR and HR cameras are first converted to color images in the ISP process. Then, distortions (\emph{e.g.}, tangential and radial distortions)  caused by cameras are reduced via camera calibration. Further, LR and HR images are aligned globally and locally. Finally, 10\% of the border is removed from each image to improve the matching accuracy of image pairs. In addition to SR, ImagePairs \cite{reza2020imagepairs} may be used for ISP and other tasks as it includes raw images. 

\begin{table}[!t]
	\center
	\caption{An overview of widely used assessment metrics for RSISR.}
	\scalebox{0.825}[0.825]{
            \begin{tabular}{m{1cm}<{\centering}m{1.5cm}<{\centering}m{2.1cm}<{\centering}m{4.2 cm}<{\centering}}
				\hline
				Metrics & Published & Full/No-reference & Keywords \\ \hline  \hline
				PSNR         &  -                                    &  Full-reference        &  Mean squared error        \\ \hline
				SSIM         & TIP-2004 \cite{wang2004image}          &  Full-reference        & Structure similarity, Luminance, Contrast, Structures         \\ \hline
				IFC         & TIP-2005 \cite{sheikh2005information}   &   Full-reference       &  Nature scene statistics, Gaussian scale mixtures        \\ \hline
				LPIPS         & CVPR-2018 \cite{zhang2018unreasonable}  &  Full-reference       & Deep features,  Human perceptual similarity        \\ \hline
				NIQE         & SPL-2012 \cite{mittal2012making}          &  No-reference        &  Quality-aware
				features, Multivariate Gaussian model         \\ \hline
				PIQE         & NCC-2015 \cite{venkatanath2015blind}      &  No-reference        &  Perceptually significant spatial regions, Block level distortion map     \\ \hline
				NRQM         & CVIU-2017 \cite{ma2017learning}           &  No-reference        &   Statistical features, Regression forests, Linear regression model      \\ \hline
			\end{tabular}}
	\label{Tab.2}
\end{table}

\subsection{Assessment Metrics for Super-Resolved Images}
\label{S.EvalMe}
In general, the quality assessment of super-resolved images is two-fold, \emph{i.e.}, human perception-based subjective evaluation and quality metrics-based objective evaluation. Overall, the former is a more direct way and it is more in agreement with the practical need. However, subjective evaluation suffers the following limitations. (\romannumeral1) The assessment result is readily affected by personal preferences.  (\romannumeral2) The evaluation process is often costly and cannot be automated. By contrast, objective evaluation is more convenient to use, although the results by different assessment metrics may not be necessarily consistent with each other as well as subjective evaluation. Table \ref{Tab.2} reports commonly used metrics for evaluating the objective quality of super-resolved images, including PSNR, SSIM \cite{wang2004image}, IFC \cite{sheikh2005information}, LPIPS \cite{zhang2018unreasonable}, NIQE \cite{mittal2012making}, PIQE \cite{venkatanath2015blind}, and NRQM \cite{ma2017learning}.  For the description, let ${\bf{X}}\in \mathbb{R} {^{H \times W \times C}}$  and ${\bf{\hat{X}}}\in \mathbb{R} {^{H \times W \times C} }$ denote the ground truth image and the super-resolved image, respectively. $H$, $W$, and $C$ denote width, height, and number of components, respectively.

\subsubsection{PSNR} Peak signal-to-noise ratio (PSNR) is the most widely used full-reference objective quality assessment metric for image restoration (\emph{e.g.}, SR, denoising, deblocking, and deblurring). Given ${\bf{\hat{X}}}$ and ${\bf{X}}$, the PSNR is defined as 
\begin{equation}\label{eq.4}
{\rm{PSNR}} = 10 \cdot {\log _{10}}(\frac{{{L^2}}}{{{\rm{MSE}}}})
\end{equation} 
where ${\rm{MSE = }}\frac{1}{{HWC}}\left\| {{\bf{X}} - {\bf{\hat X}}} \right\|_2^2$ denotes the mean squared error (MSE) between ${\bf{\hat{X}}}$ and ${\bf{X}}$, and $L$ represents the maximum pixel value (\emph{i.e.}, 255 for 8-bit images).  It can be seen from Eq.~(\ref{eq.4}) that PSNR is more concerned with the proximity between corresponding pixels in ${\bf{\hat{X}}}$ and ${\bf{X}}$, which results in the low consistency with perceptual quality in some cases.

\subsubsection{SSIM \cite{wang2004image}} The structure similarity index (SSIM) \cite{wang2004image} is a full-reference objective quality assessment metric that measures structural similarity. \footnote{The source code of SSIM \cite{wang2004image} is available at \url{https://live.ece.utexas.edu/research/Quality/index_algorithms.htm}} More specifically, the comparisons are jointly performed in the aspects of luminance, contrast, and structures as 
\begin{equation}\label{eq.5}
{\rm{SSIM = }}{\left[ {l({\bf{X}},{\bf{\hat X}})} \right]^\alpha }{\left[ {c({\bf{X}},{\bf{\hat X}})} \right]^\beta }{\left[ {s({\bf{X}},{\bf{\hat X}})} \right]^\gamma }
\end{equation} 
where $l({\bf{X}},{\bf{\hat X}}) = \frac{{2{\mu _{\bf{X}}}{\mu _{{\bf{\hat X}}}} + {C_1}}}{{{\mu _{\bf{X}}}^2 + {\mu _{{\bf{\hat X}}}}^2 + {C_1}}}$, $c({\bf{X}},{\bf{\hat X}}) = \frac{{2{\sigma _{\bf{X}}}{\sigma _{{\bf{\hat X}}}} + {C_2}}}{{{\sigma _{\bf{X}}}^2 + {\sigma _{{\bf{\hat X}}}}^2 + {C_2}}}$, and $s({\bf{X}},{\bf{\hat X}}) = \frac{{{\sigma _{{\bf{X\hat X}}}} + {C_3}}}{{{\sigma _{\bf{X}}}{\sigma _{{\bf{\hat X}}}} + {C_3}}}$. $\alpha$, $\beta$, and $\gamma$ are weighting parameters. ${{\mu _{\bf{X}}}}$ and ${{\sigma _{\bf{X}}}}$ denote the mean and standard
deviation of ${\bf{X}}$, respectively.  Similarly, ${{\mu _{\bf{\hat {X}}}}}$ and ${{\sigma _{\bf{\hat {X}}}}}$ denote the mean and standard deviation of ${\bf{\hat {X}}}$, respectively. ${{\sigma _{{\bf{X\hat X}}}}}$ is the covariance between ${\bf{\hat{X}}}$ and ${\bf{X}}$. ${C_1}$, ${C_2}$, and ${C_3}$ are constants. Further, Eq.~(\ref{eq.5}) can be 
simplified when $\alpha  = \beta  = \gamma  = 1$ and ${C_3} = \frac{{{C_2}}}{2}$ as 
\begin{equation}\label{eq.6}
{\rm{SSIM = }}\frac{{(2{\mu _{\bf{X}}}{\mu _{{\bf{\hat X}}}} + {C_1})(2{\sigma _{{\bf{X\hat X}}}} + {C_2})}}{{({\mu _{\bf{X}}}^2 + {\mu _{{\bf{\hat X}}}}^2 + {C_1})({\sigma _{\bf{X}}}^2 + {\sigma _{{\bf{\hat X}}}}^2 + {C_2})}}
\end{equation} 

In comparison, SSIM \cite{wang2004image} is reported to reflect visual quality better than PSNR. Generally, PSNR and SSIM \cite{wang2004image} are combined to assess the quality of the restored image when the corresponding ground truth image is available.

\subsubsection{IFC \cite{sheikh2005information}}  The information fidelity criterion (IFC) \cite{sheikh2005information}  is a full-reference metric that assesses the quality of images based on natural scene statistics. \footnote{The source code of IFC \cite{sheikh2005information} is available at \url{https://live.ece.utexas.edu/research/Quality/index_algorithms.htm}} Research shows that the statistics of the space formed by natural images can be characterized using various models (\emph{e.g.}, the Gaussian scale mixtures). Generally, distortions would disturb the statistics of natural scenes and make images unnatural. According to these observations, Sheikh \emph{et al.} \cite{sheikh2005information} propose to measure the visual quality of an image via jointly using the natural scene and distortion models to quantify the mutual information between the test image and reference. Overall, the IFC \cite{sheikh2005information} performs well for the quality assessment of super-resolved images \cite{yang2014single}.

\subsubsection{LPIPS \cite{zhang2018unreasonable}}  The learned perceptual image patch similarity (LPIPS) \cite{zhang2018unreasonable} is a learned metric for reference-based image quality assessment. \footnote{The source code of LPIPS \cite{zhang2018unreasonable} is available at \url{https://github.com/richzhang/PerceptualSimilarity}} More specifically, LPIPS \cite{zhang2018unreasonable} is obtained via computing the ${l_2}$ distance between the reference and the test image in a deep feature space, which shows a good agreement with human judgments.

\subsubsection{NIQE \cite{mittal2012making}}  The natural image quality evaluator (NIQE) is a completely blind metric without the knowledge of human judgments or distortions \cite{mittal2012making}. \footnote{The source code of NIQE \cite{mittal2012making} is available at \url{https://live.ece.utexas.edu/research/Quality/index_algorithms.htm}} The multivariate Gaussian (MVG) model is used to fit the ``quality-aware" features extracted from images. More specifically, the features include the parameters of the generalized Gaussian distribution (GGD) and the asymmetric generalized Gaussian distribution (AGGD) that characterize the behavior of image patches. Then, the quality of an image is measured using the distance between the two MVG models fitting natural images and the evaluated image. 

\subsubsection{PIQE \cite{venkatanath2015blind}} The perception-based quality evaluator (PIQE) is a no-reference image quality assessment metric \cite{venkatanath2015blind}. \footnote{The implementation of PIQE \cite{venkatanath2015blind} is included in Matlab as \url{https://www.mathworks.com/help/images/ref/piqe.html}} Considering that the attention of human visual system (HVS) is highly directed towards spatially active regions, the test image is divided into non-overlapping blocks and block-level analysis is conducted to identify distortion and grade quality. Therefore, PIQE \cite{venkatanath2015blind} can provide a spatial quality map. The overall quality of the evaluated image can be obtained by pooling the block level quality scores. 

\subsubsection{NRQM \cite{ma2017learning}} This is a learned no-reference quality metric (NRQM) for assessing super-resolved images \cite{ma2017learning}. \footnote{The source code of NRQM \cite{ma2017learning} is available at \url{https://github.com/chaoma99/sr-metric}}  To predict the perceptual scores of super-resolved images, three groups of statistical features including local frequency features, global frequency features, and spatial features are extracted. The selected features cover the distribution of discrete cosine transform coefficients, the distribution of wavelet coefficients, the spatial discontinuity property of pixel intensity, \emph{etc}. On this basis, three regression forests are used to model these features independently and their results are combined linearly to estimate the final perceptual score. These three forests and the linear regression model are trained on a large-scale dataset of super-resolved images with perceptual scores. Overall, the visual quality predicted by NRQM \cite{ma2017learning} matches subjective evaluation well on SR results.

\begin{figure}[!tb]
	\centering
	\includegraphics[width = 6 cm]{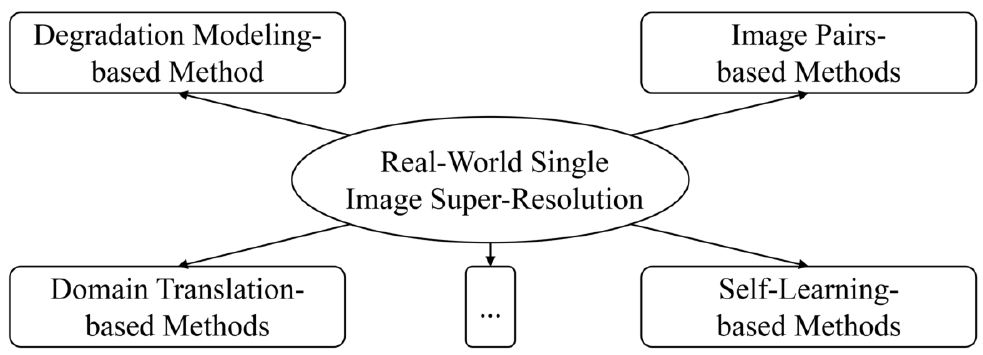}
	\caption{The taxonomy of existing real-world single image super-resolution techniques.}
	\label{Fig.2}
\end{figure}

\section{Technologies and Methods}
\label{S.TechMe}
\begin{figure*}[!t]
	\centering
	\includegraphics[width = 18 cm]{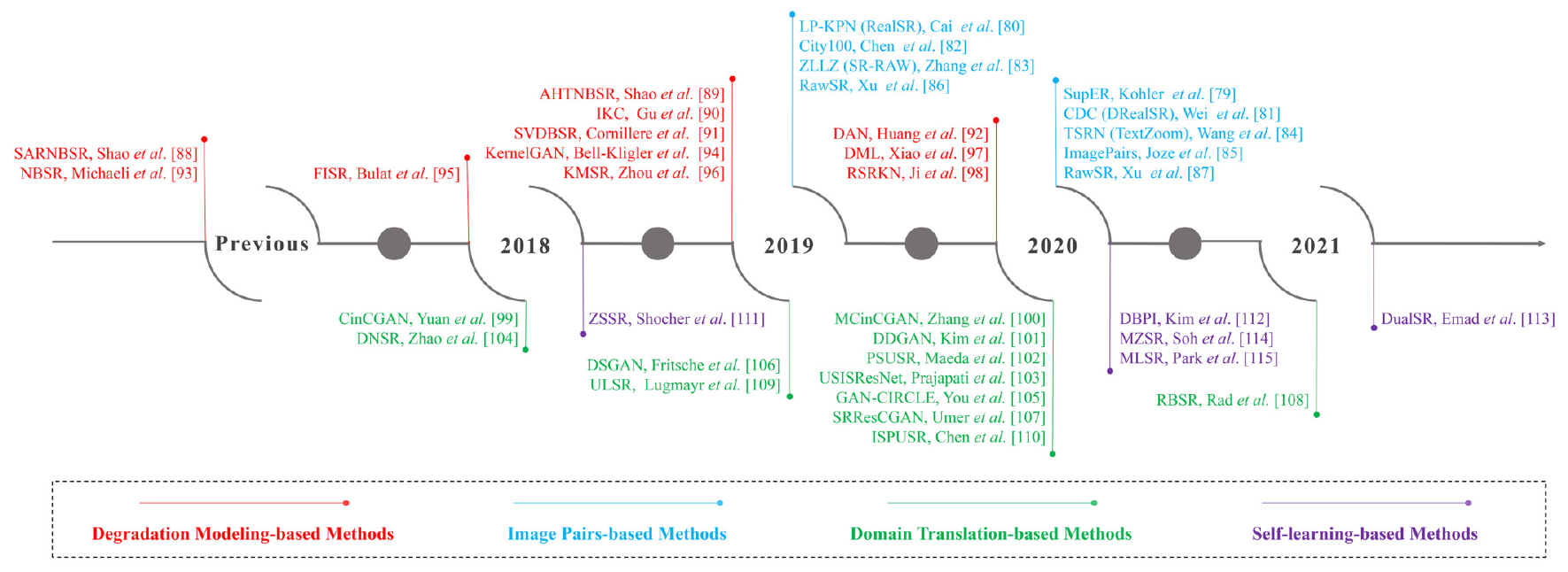}
	\caption{Milestones of RSISR methods.}
	\label{Fig.3}
\end{figure*}

Researchers have been studying the SISR methods for practical applications. Especially with the SR performance on synthetic data becoming better and better, more and more attention has been paid to RSISR. Fig. \ref{Fig.2} presents the overall taxonomy of existing RSISR techniques. Note that we focus more on deep learning-based methods. According to the  primary principles and characteristics of existing RSISR methods, we group them into four categories, \emph{i.e.}, degradation modeling-based methods \cite{shao2015simple, shao2019nonparametric, gu2019blind, cornillere2019blind, huang2020unfolding, michaeli2013nonparametric, bell2019blind, bulat2018learn, zhou2019kernel, xiao2020degradation, ji2020real}, image pairs-based methods \cite{cai2019toward, zhang2019zoom, chen2019camera, wang2020scene, kohler2019toward, wei2020component, reza2020imagepairs, xu2019towards, xu2020eploiting}, domain translation-based methods \cite{yuan2018unsupervised, zhang2019multiple, kim2020unsupervised, maeda2020unpaired, prajapati2020unsupervised, zhao2018unsupervised, you2019ct, fritsche2019frequency, muhammad2020deep, rad2021benefiting, lugmayr2019unsupervised, chen2020unsupervised}, and self-learning-based methods \cite{shocher2018zero, bell2019blind, kim2020dual, emad2021dualsr, soh2020meta, park2020fast}. Fig. \ref{Fig.3} and Table \ref{Tab.3} summarize existing RSISR methods. It is worth noting that one method may belong to different categories when it is viewed from different perspectives. The following sections introduce these methods in detail.

\begin{table*}[!t]
	\center
	\caption{An overview of existing works on RSISR.}
	\scalebox{0.9}[0.9]{
		\begin{tabular}{m{2.5cm}<{\centering}m{3cm}<{\centering}m{4cm}<{\centering}m{7cm}<{\centering}}
			\hline 
			Methods & Published & Category & Keywords \\ \hline  \hline
			
			SupER &   TPAMI-2020 \cite{kohler2019toward}       &   Image pairs-based       &   Benchmarking super-resolution on real data      \\ \hline
			LP-KPN (RealSR) &    ICCV-2019 \cite{cai2019toward}       &   Image pairs-based       &  Laplacian pyramid-based kernel prediction network         \\ \hline
			CDC (DRealSR) &  ECCV-2020  \cite{wei2020component}       &   Image pairs-based       &  Component divide-and-conquer model,  Gradient weighted loss         \\ \hline
			CameraSR (City100) &  CVPR-2019 \cite{chen2019camera}       &   Image pairs-based       &   SR from the perspective of camera lenses        \\ \hline
			ZLLZ (SR-RAW) &    CVPR-2019 \cite{zhang2019zoom}       &   Image pairs-based       &    RAW sensor data, Contextual bilateral loss     \\ \hline
			TSRN (TextZoom) &  ECCV-2020  \cite{wang2020scene}       &   Image pairs-based       &   Sequential residual block,  Central alignment module, Gradient profile loss       \\ \hline
			ImagePairs &    CVPRW-2020 \cite{reza2020imagepairs}       &   Image pairs-based       &   A new data acquisition
			technique for gathering real image data       \\ \hline
			RawSR &    CVPR-2019 \cite{xu2019towards},  TPAMI-2020 \cite{xu2020eploiting}       &   Image pairs-based       &  Two-branch structure, Raw image, Color correction        \\ \hline  \hline 
			
			SARNBSR & ICIG-2015 \cite{shao2015simple}          &    Degradation modeling-based      &   Blur kernel estimation, Bi-$l_0$-$l_2$-norm regularization       \\ \hline
			AHTNBSR & JMIV-2019 \cite{shao2019nonparametric}   &    Degradation modeling-based       &   Adaptive heavy-tailed image priors, Nonparametric blur kernel estimation       \\ \hline
			IKC     & CVPR-2019  \cite{gu2019blind}         &    Degradation modeling-based       &  Iterative kernel correction, Spatial feature transform         \\ \hline
			SVDBSR  & TOG-2019 \cite{cornillere2019blind}    &    Degradation modeling-based     &  Degradation-aware SR network,  Kernel discriminator network      \\ \hline
			DAN     & NeurIPS-2020 \cite{huang2020unfolding}     &    Degradation modeling-based       &  SR image restorer, Blur kernel estimator,  End-to-end trainable        \\ \hline
			NBSR    &  ICCV-2013 \cite{michaeli2013nonparametric}         &  Degradation modeling-based        &    Kernel estimation, Recurrence of small image patches    \\ \hline
			KernelGAN &  NeurIPS-2019 \cite{bell2019blind}         &   Degradation modeling-based      &   Deep internal learning, Cross-scale recurrence property, GAN, Image-specific SR-kernel     \\ \hline
			FISR    &   ECCV-2018 \cite{bulat2018learn}        &   Degradation modeling-based       &  High-to-low GAN, Low-to-high GAN, Unpaired LR and HR images      \\ \hline
		    KMSR    &   ICCV-2019 \cite{zhou2019kernel}        &   Degradation modeling-based      &   Realistic blur kernels, Blur kernel pool augment, GAN     \\ \hline
			DML     &  ACCV-2020 \cite{xiao2020degradation}     &    Degradation modeling-based       &   Realistic HR-LR image pair synthesis,  Pixel-wise spatially variant degradation kernel       \\ \hline
			RSRKN   &  CVPRW-2020  \cite{ji2020real}       &   Degradation modeling-based       &   Degradation framework, Kernel estimation, Noise injection, GAN        \\ \hline \hline

            CinCGAN &  CVPRW-2018  \cite{yuan2018unsupervised}       &  Domain translation-based        &  Unsupervised learning, Cycle-in-cycle network structure, GAN        \\ \hline
            MCinCGAN &  TIP-2020  \cite{zhang2019multiple}       &  Domain translation-based        &    Unsupervised learning, Multiple cycle-in-cycle network structure, GAN      \\ \hline
            DDGAN &  CVPRW-2020  \cite{kim2020unsupervised}       &  Domain translation-based        &   Unsupervised learning, Unknown degradation, Cycle-in-cycle GAN, Domain discriminator       \\ \hline
            PSUSR &   CVPR-2020 \cite{maeda2020unpaired}       &  Domain translation-based        &   Unpaired kernel/noise correction network,  Pseudo-paired SR network, GAN      \\ \hline
            USISResNet &   CVPRW-2020 \cite{prajapati2020unsupervised}       &  Domain translation-based        &  Unsupervised learning, GAN, Mean Opinion Score-based loss function   \\ \hline
            DNSR &   arXiv-2018 \cite{zhao2018unsupervised}       &  Domain translation-based        &    Unsupervised degradation learning, Bidirectional structural consistency,  Bi-cycle network     \\ \hline
            GAN-CIRCLE &  TMI-2020   \cite{you2019ct}       &  Domain translation-based        &   Computed tomography, CycleGAN       \\ \hline
            DSGAN &   ICCVW-2019 \cite{fritsche2019frequency}       &  Domain translation-based        &    Frequency separation,  Unsupervised learning,  GAN     \\ \hline
            SRResCGAN &    CVPRW-2020  \cite{muhammad2020deep}       &  Domain translation-based        &    Image observation model, Domain learning, GAN      \\ \hline
            RBSR &   WACV-2021 \cite{rad2021benefiting}       &  Domain translation-based        &   Bicubically down-sampled
            images, GAN      \\ \hline
            ULSR &    ICCVW-2019 \cite{lugmayr2019unsupervised}       &  Domain translation-based        &    Unsupervised learning, Unpaired data, GAN      \\ \hline
            ISPUSR &  CVPRW-2020 \cite{chen2020unsupervised}       &  Domain translation-based        &   Unsupervised image translation, Supervised SR, Collaborative training \\ \hline \hline
            
           ZSSR  &    CVPR-2018 \cite{shocher2018zero}       &  Self-learning-based        &    Zero-shot, Internal recurrence, Deep internal learning, Image-specific CNN     \\ \hline
           DBPI &    SPL-2020 \cite{kim2020dual}       &   Self-learning-based       &    Unified internal learning,  Downscaling/SR network, Dual back-projection loss    \\ \hline
           DualSR &  WACV-2021  \cite{emad2021dualsr}       &   Self-learning-based       &   Zero-shot,  Dual-path architecture, GAN, Masked interpolation loss     \\ \hline
           MZSR &    CVPR-2020 \cite{soh2020meta}       &   Self-learning-based       &   Meta-transfer learning,  Large-scale training, External and internal information     \\ \hline
           MLSR &    ECCV-2020 \cite{park2020fast}       &   Self-learning-based       &   Meta-learning, Patch-recurrence property       \\ \hline 
	\end{tabular}}
	\label{Tab.3}
\end{table*}

\subsection{Degradation Modeling-based Methods}
\label{S.DegModel}
\begin{figure}[!tb]
	\centering
	\includegraphics[width = 5.5 cm]{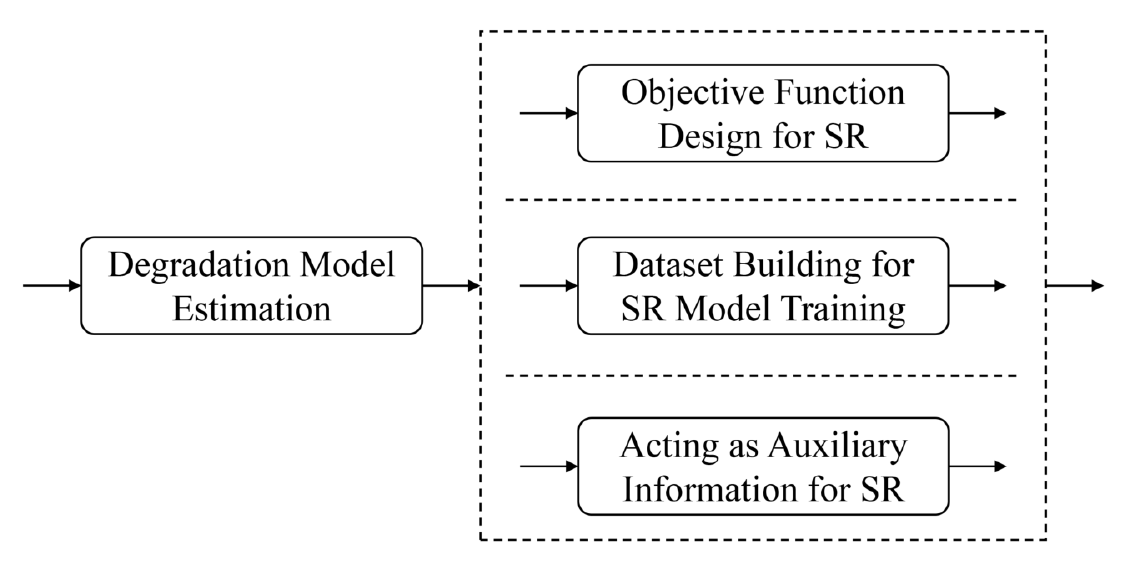}
	\caption{The general idea of degradation modeling-based methods.}
	\label{Fig.4}
\end{figure}

Compared with the SR of synthetic LR images, one of the main challenges for RSISR is that the degradation model (\emph{i.e.}, the $D(\cdot)$ in Eq.~(\ref{eq.1})) is unknown. Generally, the degradation parameter set is necessary to 
design the objective function for reconstruction-based SISR methods derived from the \emph{Maximum a Posteriori} (MAP) estimation. For learning-based methods that aim to obtain the mapping from LR images to their HR  counterparts, the degradation parameter set is crucial for building training datasets. Therefore, as presented in Fig. \ref{Fig.4}, an intuitive way is to estimate the degradation parameter of the LR input prior to super-resolving or to iteratively optimize the degradation parameter and the super-resolved image. Eq.~(\ref{eq.3})) shows the commonly used degradation model, in which the blur kernel $\bf{B}$ and noise $\bf{n}$ are unknowns. The exploration by Efrat \emph{et al.} \cite{efrat2013accurate} shows that an accurate blur model is more important to the success of SR than a sophisticated image prior, in particular for real-world images. As highlighted in \cite{efrat2013accurate}, most of the existing degradation modeling-based methods focus on the recovery of the blur kernel. \footnote{In some cases, the noise is not explicitly characterized. Natural image priors are generally incorporated to suppress noise.} 

RSISR is a severely ill-posed problem. Shao \emph{et al.} \cite{shao2015simple} formulate this kind of task as an energy minimization problem and jointly optimize the nonparametric blur kernel and the intermediate super-resolved image. Mathematically, the objective function for blur kernel estimation is defined as
\begin{equation}\label{eq.7}
\mathop {\min }\limits_{{\bf{X}},{\bf{b}}} {\kern 2pt} \lambda \left\| {{\bf{SBX}} - {\bf{Y}}} \right\|_2^2 + \Re ({\bf{X}},{\bf{b}}) + \eta \left\| {{\bf{BX}} - {\bf{\tilde X}}} \right\|_2^2
\end{equation} 
where $\lambda $ and $\eta$ are parameters for balancing different terms. ${\bf{\tilde X}}$ denotes the super-resolved image generated by a non-blind learning-based method, and $\bf{b}$ denotes the blur kernel corresponding to the blur matrix $\bf{B}$. $\Re ({\bf{X}},{\bf{b}})$ represents the direct bi-${l_0}$-${l_2}$-norm regularization term for the intermediate super-resolved image $\bf{X}$ and the blur kernel $\bf{b}$, which is beneficial to the accurate estimation of $\bf{b}$. After iteratively optimizing the $\bf{X}$ and $\bf{b}$ in Eq. \ref{eq.7}, the estimated blur kernel $\bf{\hat{b}}$ can be combined with non-blind SR methods to produce an HR estimate. In \cite{shao2019nonparametric}, Shao \emph{et al.} introduce the ${l_{\alpha}}$-norm-based adaptive heavy-tailed image prior to further improve the above approach. The studies in \cite{shao2015simple} and \cite{shao2019nonparametric} demonstrate that, with the aid of effective constraints, the iterative optimization of the blur kernel and the super-resolved image is good for the accuracy of blur kernel estimation. 

Different from the above numerical optimization-based approaches \cite{shao2015simple, shao2019nonparametric}, the blur kernel and the super-resolved image can also be jointly optimized using deep neural networks \cite{gu2019blind, cornillere2019blind, huang2020unfolding}. Generally, the mismatch of blur kernels would result in artifacts in the super-resolved image, \emph{e.g.}, over-smoothing or over-sharpening. Based on this observation, Gu \emph{et al.} \cite{gu2019blind} and Cornillere \emph{et al.} \cite{cornillere2019blind} propose to progressively correct the inaccurate kernel according to the quality of the super-resolved image. \footnote{The source code of IKC \cite{gu2019blind} is available at \url{https://github.com/yuanjunchai/IKC}} More specifically, they develop degradation-aware SR networks to produce HR images, in which the blur kernel is utilized as auxiliary information for SR. Meanwhile, corresponding deep neural networks are designed to correct the kernel with the guidance of the intermediate SR result. Unlike previous approaches \cite{gu2019blind, cornillere2019blind} that combine two or more networks, Huang \emph{et al.} \cite{huang2020unfolding} develop a deep alternating network (DAN) for RSISR, in which the iterative optimization process between the super-resolved image and the blur kernel is unfolded to an end-to-end trainable network. \footnote{The source code of DAN \cite{huang2020unfolding} is available at \url{https://github.com/greatlog/DAN}} DAN \cite{huang2020unfolding} consists of a chain of alternately stacked restorers and estimators, which are responsible for the restoration of the HR image and the estimation of the blur kernel, respectively. For the above methods, the super-resolved image and the blur kernel corresponding to the LR input are iteratively refined. After joint optimization, the refined blur kernel and the super-resolved image are supposed to be more accurate.

Previous studies have shown that natural image priors such as patch recurrence property are useful for degradation modeling. In \cite{michaeli2013nonparametric}, Michaeli \emph{et al.} point out that the Point Spread Function (PSF) is not the optimal blur kernel, and they further propose to obtain the principled MAP estimate of the blur kernel via maximizing the similarity of recurring image patches across scales of the LR input. \footnote{The project homepage of NBSR \cite{michaeli2013nonparametric} is at \url{http://www.wisdom.weizmann.ac.il/~vision/BlindSR.html}} The estimated blur kernel can be used to degrade the LR input or natural HR images artificially. In this way, the blur kernel estimation approach is smoothly plugged into both self-example-based and external-example-based SR approaches \cite{glasner2009super, zeyde2010single}. The results in \cite{michaeli2013nonparametric} show that the accuracy enhancement of blur kernel estimation leads to an obvious improvement of SR performance on synthetic as well as real-world images. 

Degradation modeling is also vitally important for deep learning-based SR approaches. Deep convolutional neural networks (CNN)-based SISR approaches usually achieve state-of-the-art (SOTA) results on standard benchmarks. Nevertheless, their performance is limited when applied to real-world images. The main reason is that the kernel (\emph{e.g.}, ``bicubic'' kernel) used to generate training data is significantly different from the blur in a real scenario. To address this problem, some recently presented deep learning-based RSISR methods \cite{bell2019blind, bulat2018learn, zhou2019kernel, xiao2020degradation, ji2020real} adopt the pre-estimated degradation parameters to generate samples for model training. For example, inspired by \cite{michaeli2013nonparametric}, Bell-Kligler  \emph{et al.} \cite{bell2019blind} develop an image-specific internal-GAN (\emph{i.e.}, KernelGAN) to learn the internal distribution of patches. \footnote{The source code of KernelGAN \cite{bell2019blind} is available at \url{https://github.com/sefibk/KernelGAN} \label{FNKernelGAN} } The KernelGAN \cite{bell2019blind} is trained solely using the LR test image, making its discriminator unable to differentiate the patch distribution of the original LR input from that of the degraded version of the LR image produced by the generator. After the joint training with the discriminator, the generator can well characterize the degradation process with an image-specific kernel. Then, the LR test image and its degraded version generated by the generator form paired data for SR model training. Bulat \emph{et al.} \cite{bulat2018learn} train a generative adversarial network (GAN)-based degradation model from unpaired HR and LR face images, and then use the learned network to generated image pairs for SR network training. \footnote{The source code of FISR \cite{bulat2018learn} is available at \url{https://github.com/jingyang2017/Face-and-Image-super-resolution}} Zhou \emph{et al.} \cite{zhou2019kernel} propose to obtain a group of realistic blur kernels from real-world photographs and a GAN is trained on them to augment the pool of realistic blur kernels. \footnote{The source code of KMSR \cite{zhou2019kernel} is available at \url{https://github.com/IVRL/Kernel-Modeling-Super-Resolution}} With the augmented kernel pool, more realistic and diverse LR-HR image pairs can be constructed to train the SR model. Analogously, Xiao \emph{et al.} \cite{xiao2020degradation} model spatially variant degradation via learning a set of basis blur kernels and corresponding pixel-wise weights from real-world image pairs. The learned realistic degradation model is then used to generate pseudo-realistic LR-HR image pairs. More recently, Ji \emph{et al.} \cite{ji2020real} take this idea one step further and develop an effective degradation framework using various realistic blur kernels and noise distributions, winning the \emph{NTIRE 2020 Challenge on Real-World Image Super-Resolution} \cite{lugmayr2020ntire}. \footnote{The source code of RSRKN \cite{ji2020real} is available at \url{https://github.com/jixiaozhong/RealSR}} The outstanding performance of degradation modeling-based RSISR methods demonstrates that degradation modeling is meaningful and this kind of approach is a feasible solution to the SR of real-world images.
\begin{figure}[!tb]
	\centering
	\includegraphics[width = 6 cm]{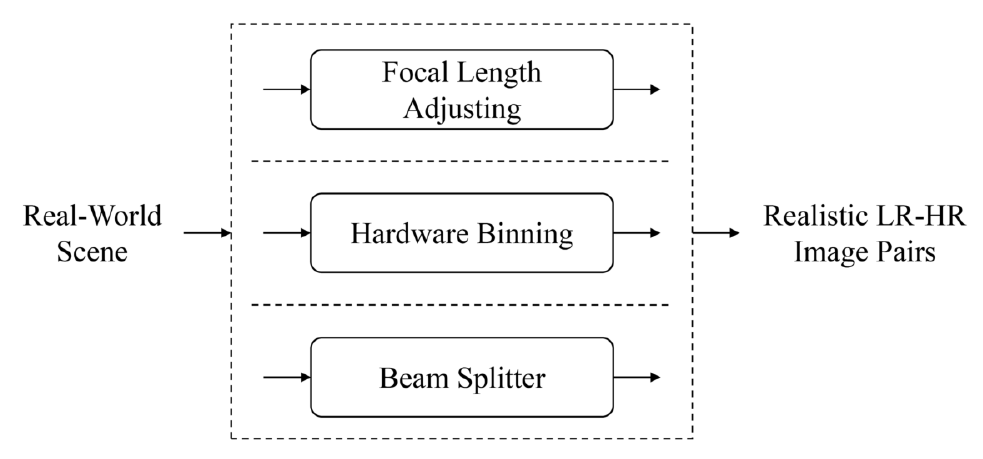}
	\caption{Existing solutions to construct realistic LR-HR image pairs.}
	\label{Fig.5}
\end{figure}

\subsection{Image Pairs-based Methods}
\label{S.ImaPair}
Although paired LR-HR training data can be synthesized from high-quality images according to pre-defined degradation models \cite{zhou2019kernel, ji2020real}, deriving explicit realistic degradation models from real-world images is challenging. To deal with this problem, researchers \cite{cai2019toward, zhang2019zoom, chen2019camera, wang2020scene, kohler2019toward, wei2020component, reza2020imagepairs} propose to directly collect the images of the same scene with different resolutions, which are used to construct realistic LR-HR image pairs for RSISR model training. Overall, as illustrated in Fig. \ref{Fig.5}, currently there are three main ways to collect real-world images for dataset building, including the focal length adjusting-based approach \cite{cai2019toward, zhang2019zoom, chen2019camera, wang2020scene, wei2020component}, the hardware binning-based approach \cite{kohler2019toward}, and the beam splitter-based approach \cite{reza2020imagepairs}. The representative realistic datasets built with the above image collection approaches are described in Section \ref{S.DataSet}. Therefore, this section focuses on the RSISR models developed on these real-world datasets. Intuitively, given LR-HR image pairs, nearly all existing supervised SR methods (\emph{e.g.}, SRCNN \cite{dong2015image}, VDSR \cite{kim2016accurate}, EDSR \cite{lim2017enhanced}, RDN \cite{zhang2018residual}, SRGAN \cite{ledig2017photo}, \emph{etc}.) can be adopted to learn the mapping from LR images to their HR counterparts. The mapping learned from realistic datasets is supposed to apply to the SR of real-world images. However, in fact challenges remain. 

For instance, the degradation kernels of real-world images are generally non-uniform, varying with the depth in a real scene. Thus, training an SR model that treats all pixels the same way as most previous deep CNN-based SR approaches may not be the optimal solution. For this problem, Cai  \emph{et al.} \cite{cai2019toward} propose the LP-KPN, which combines the Laplacian pyramid with the pixel-wise kernel prediction network (KPN), achieving good SR performance and high efficiency. \footnote{The source code of LP-KPN \cite{cai2019toward} is available at \url{https://github.com/csjcai/RealSR}}  Another challenge that cannot be ignored is the misalignment between LR and HR image pairs in the collected realistic datasets. Although image registration is performed to align realistic image pairs, misalignment is unavoidable. As a consequence of misalignment, blurring artifacts may be introduced in the reconstructed HR images when using these datasets to train the SR model with pixel-to-pixel losses (\emph{e.g.}, $l_1$ and $l_2$). Inspired by the Contextual Loss  \cite{mechrez2018contextual} and the edge-preserving bilateral filter \cite{tomasi1998bilateral}, Zhang  \emph{et al.} \cite{zhang2019zoom} propose the Contextual Bilateral loss (CoBi)  to resolve this issue. \footnote{The source code of ZLLZ \cite{zhang2019zoom} is available at \url{https://github.com/ceciliavision/zoom-learn-zoom}}  CoBi integrates the pixel-level information and spatial pixel coordinates to measure image similarity. Moreover, two spaces including RGB image patches and pre-trained perceptual-features (\emph{e.g.}, VGG-19 \cite{simonyan2015very}) are jointly considered in CoBi to improve performance further. It is demonstrated that CoBi is robust to the mild misalignment in the realistic image pairs for supervised SR model training. Considering that pixel-wise losses are generally more focused on smoothing flat regions and sharping edges while neglecting the recovery of realistic details of textures to some extent, Wei \emph{et al.} \cite{wei2020component} develop a Component Divide-and-Conquer (CDC) SR model for real-world images.  \footnote{The source code of CDC \cite{wei2020component} is available at \url{https://github.com/xiezw5/Component-Divide-and-Conquer-for-Real-World-Image-Super-Resolution}} More specifically, the flat, edge, and corner components are first predicted by three component-attentive blocks respectively in CDC and then they are aggregated to produce the final SR image based on the learned component-attentive maps. To achieve this goal, a gradient-weighted loss that can adapt the model training to the reconstruction difficulties of different image components is applied. The results on RealSR \cite{cai2019toward} and DRealSR \cite{wei2020component}  prove the superiority and  generalization capability of CDC \cite{wei2020component}. Given a dataset containing raw and color images, learning a mapping from LR raw images to HR color images using deep neural networks is an intuitive approach to exploit raw images for SR. However, one raw image could correspond to a set of color images because it does not have the information for the processes (\emph{e.g.}, color correction) within the image signal processing system, making the above naive way do not work well. To address this problem, Xu \emph{et al.} \cite{xu2019towards, xu2020eploiting} design a two-branch CNN which jointly exploits the LR raw data and corresponding LR color image to recover fine structures and high-fidelity color appearances.

In addition to the above challenges shared by almost all real-world images, the SR of certain kinds of images (\emph{e.g.}, text images, remote sensing images, and medical images) usually has a particularity. Therefore, specific SR models should be designed for these scenarios. For example, the TSRN developed by Wang \emph{et al.} \cite{wang2020scene} is an SR network for real scene text images. \footnote{The source code of TSRN \cite{wang2020scene} is available at \url{https://github.com/JasonBoy1/TextZoom}} To leverage the strong sequential characteristics of text images, the Bi-directional LSTM (BLSTM) mechanism is added to basic residual blocks. In order to address the misalignment problem in the realistic text image dataset TextZoom, TSRN \cite{wang2020scene} introduces a spatial transform network-based central alignment module in the front of the network. Furthermore, aiming at enhancing the shape boundary of characters, a gradient prior loss is combined with the $l_2$ loss to train TSRN. It is demonstrated that the SR of real-world text images using TSRN does increase the recognition accuracy. Predictably, in some cases, SR can also benefit other computer vision  tasks, \emph{e.g.}, objection detection \cite{haris2018task} and semantic segmentation \cite{wang2020dual}. 

\subsection{Domain Translation-based Methods}
\label{S.DomTrans}
\begin{figure}[!tb]
	\centering
	\includegraphics[width = 9 cm]{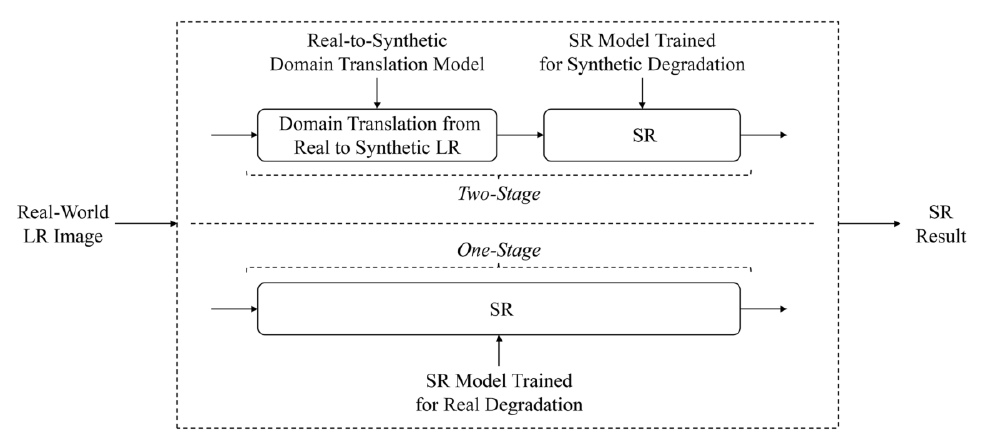}
	\caption{The general idea of domain translation-based methods. }
	\label{Fig.6}
\end{figure}

As introduced in Sections \ref{S.DataSet} and \ref{S.ImaPair}, it is hard to obtain real-world datasets with well-aligned LR-HR image pairs. More often than not, what we have are only LR images for model training in practical applications. Or better yet, a set of HR images are available for reference besides LR training images, but there is no one-to-one correspondence between LR and HR images. The supervised approaches no longer apply in these cases due to the lack of paired samples. Previous studies \cite{yuan2018unsupervised, zhang2019multiple, kim2020unsupervised, maeda2020unpaired, prajapati2020unsupervised, zhao2018unsupervised, you2019ct, fritsche2019frequency, muhammad2020deep, muhammad2020deep, rad2021benefiting, lugmayr2019unsupervised, chen2020unsupervised} demonstrate that domain translation is a feasible solution to deal with this problem. For this kind of RSISR approach, real-world LR images, synthetic LR images (also known as clean LR images or ideal LR images),  and HR images are thought to be in different domains. Consequently, the SR of real-world images converts to the translation from the real-world LR image domain (RLRD) to the HR image domain (HRD). Overall, as presented in Fig. \ref{Fig.6}, there are two main ways to cross RLRD to HRD, \emph{i.e.}, the two-stage and the one-stage approaches, and the most notable difference lies in whether the synthetic LR image domain (SLRD) is used as a relay station.

Most existing SISR approaches are trained using synthetic data, thereby achieving excellent performance on clean LR images. As is known to all, there is a noticeable domain gap between SLRD and RLRD, causing the degradation of SR accuracy on real-world images. Intuitively, one can mitigate this performance degradation via reducing the domain gap, which is how the two-stage domain translation-based RSISR approaches work. As shown in Fig. \ref{Fig.6}, the two-stage approaches generally include two main steps, \emph{i.e.}, domain translation and SR. Usually, generic SR methods are adopted in the SR reconstruction phase, so as to benefit from pre-trained SR models learned from large-scale datasets. The Cycle-in-Cycle GAN (CinCGAN) developed by Yuan \emph{et al.} \cite{yuan2018unsupervised} is a representative of this kind of RSISR method. More specifically, CinCGAN \cite{yuan2018unsupervised} first uses a domain translation network to map realistic LR images in RLRD into the SLRD, and then a pre-trained deep SR network with the ideal degradation assumption is stacked to upscale the translation result to the desired size. Finally, the domain translation and SR networks are end-to-end fine-tuned. Inspired by CycleGAN \cite{zhu2017unpaired}, the domain translation network in CinCGAN \cite{yuan2018unsupervised} is trained to map realistic LR inputs to synthetic LR images (\emph{i.e.}, ``bicubic''-downsampled images) using unpaired training data, which usually can suppress the artifacts such as noise in real-world LR images to make them more suitable for the following SR network. Results show that the CinCGAN \cite{yuan2018unsupervised} trained with unpaired data achieves comparable SR performance as supervised methods. In MCinCGAN \cite{zhang2019multiple}, the CinCGAN \cite{yuan2018unsupervised} is improved via introducing a progressive multi-cycle framework for large-scale upsampling and a new constraint to suppress the color fluctuation in training. For more stable model training and better SR performance, the DDGAN developed by Kim \emph{et al.} \cite{kim2020unsupervised} further combines the pixel-wise loss, the VGG feature loss, and the SSIM loss to measure similarity. \footnote{The source code of DDGAN \cite{kim2020unsupervised} is available at \url{https://github.com/GT-KIM/unsupervised-super-resolution-domain-discriminator}}  Meanwhile, a domain discriminator that takes the noise, texture, and color into consideration simultaneously is proposed to make the generated image more consistent with the target domain distribution. More recently, Maeda \emph{et al.} \cite{maeda2020unpaired} propose an end-to-end trainable framework UISRPS to jointly optimize the domain translation network and the SR network, achieving excellent SR results on real-world face images and aerial images. Benefiting the architecture, it is convenient to integrate existing SR networks and pixel-wise loss functions into UISRPS \cite{maeda2020unpaired}. 

Different from the two-stage methods mentioned above, the one-stage domain translation-based RSISR frameworks aim to produce a super-resolved image directly from the real-world LR input, as presented in Fig. \ref{Fig.6}. How to learn the translation mapping from RLRD to HRD without real LR-HR image pairs is the crucial issue. Prajapati \emph{et al.} \cite{prajapati2020unsupervised} propose to train a GAN-based network USISRNet to upsample real-world images. \footnote{The source code of USISResNet \cite{prajapati2020unsupervised} is available at \url{https://github.com/kalpeshjp89/USISResNet}} Since only unpaired LR and HR training images are available, USISRNet is optimized through unsupervised learning. Beyond the standard GAN losses, a pixel-wise content loss, a Total-Variation loss, and a quality assessment loss are combined to optimize USISRNet. The content loss makes the SR result be not far from the bicubically upsampled version, thus preserving the primary content of the LR image. The Total-Variation loss is integrated to suppress the noise and artifacts. For better perceptual quality of the super-resolved image, the learned Mean Opinion Score is used to construct the quality assessment loss. Thanks to the combination of multiple losses, USISRNet \cite{prajapati2020unsupervised} achieves good generalization capacity. Inspired by CycleGAN \cite{zhu2017unpaired}, some researchers propose to learn the direct relationship between RLRD to HRD using cycle consistency-constrained GANs  \cite{zhao2018unsupervised, you2019ct}. \footnote{The source code of GAN-CIRCLE \cite{you2019ct} is available at \url{https://github.com/charlesyou999648/GAN-CIRCLE}} Given an LR image, the LR-to-HR generator (\emph{i.e.}, RLRD to HRD) is trained to reconstruct a vivid HR image that can be returned to the LR input by the corresponding HR-to-LR generator (\emph{i.e.}, HRD to RLRD). Conversely, an HR image should be well recovered by the LR-to-HR generator from its downsampled version produced by the HR-to-LR generator. After the joint training of the two generators and corresponding discriminators, the LR-to-HR generator models the direct mapping from RLRD to HRD and it is used to reconstruct HR images from realistic LR images. 

Considering that learning the end-to-end translation between RLRD to HRD from unpaired LR-HR data is challenging, part of the one-stage domain translation-based RSISR methods \cite{lugmayr2019unsupervised, fritsche2019frequency, rad2021benefiting, chen2020unsupervised, muhammad2020deep}  also use the synthetic LR image as a bridge in the training phase. Given unpaired realistic LR and HR images, Fritsche \emph{et al.} \cite{fritsche2019frequency} first bicubically downsample HR images.  Bicubically downsampled results are then translated into the realistic domain to make them follow real scene characteristics, using a standard GAN-based domain translation network trained on the bicubically downsampled images and realistic LR images in an unsupervised fashion. Taking the pseudo-realistic LR images and corresponding HR images as training sample pairs, the ESRGAN \cite{wang2018esrgan} is trained for upsampling in a supervised manner. In order to generate images well matching the target distribution, both domain translation and SR networks are optimized with frequency separation-based loss functions. \footnote{The source code of DSGAN \cite{fritsche2019frequency} is available at \url{https://github.com/ManuelFritsche/real-world-sr}} More specifically, the color loss, the texture loss, and the perceptual loss are employed to the low-frequency component, the high-frequency component, and the whole image, respectively.  Note that only the SR model is needed to upscale real-world images in the testing phase because it is trained on the image pairs that follow real-world image distribution. On this basis, recently Umer \emph{et al.} \cite{muhammad2020deep} improve the GAN-based SR model following the real-world image observation model, thus exploiting the powerful regularization and optimization techniques simultaneously. \footnote{The source code of SRResCGAN \cite{muhammad2020deep} is available at \url{https://github.com/RaoUmer/SRResCGAN}}   Rad \emph{et al.} \cite{rad2021benefiting} convert realistic LR images to bicubic look-alike images based on their copying mechanism and bicubic perceptual loss. Different from the above works \cite{fritsche2019frequency, muhammad2020deep, rad2021benefiting} which use a single direction domain translation model, Lugmayr \emph{et al.} \cite{lugmayr2019unsupervised} and Chen \emph{et al.}\cite{chen2020unsupervised} propose to train a bi-directional domain translation model with the cycle consistency constraints for better robustness.  

Overall, domain translation is an effective way to reduce the domain gap between synthetic and realistic data, thus improving the generalization capability of SR models on ever-changing real-world images. In contrast, the two-stage domain translation-based RSISR approaches can integrate synthetic data-trained SR models more elegantly, while the one-stage methods generally have lower complexity in the testing phase.

\begin{figure}[!tb]
	\centering
	\includegraphics[width = 5.2 cm]{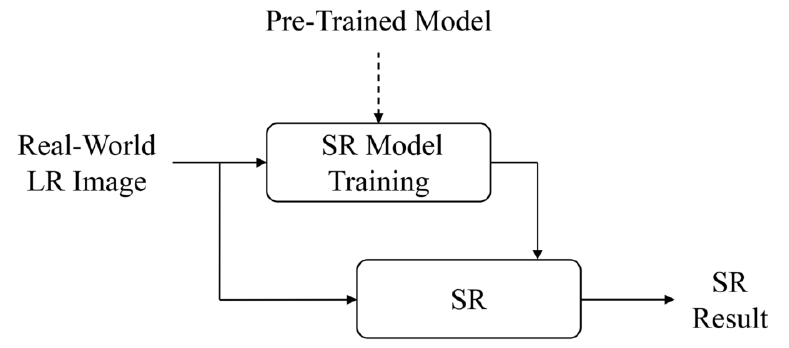}
	\caption{The general idea of self-learning-based SR methods. Note that the pre-trained model is optional.}
	\label{Fig.7}
\end{figure}
\subsection{Self-Learning-based Methods}
\label{S.SelfLearn}
Most existing RSISR methods use external dataset (\emph{i.e.}, paired or unpaired training data) to train SR models. Therefore, the SR performance is tightly bound to the consistency between testing data and training data. However, real-world images do not always obey the characteristics of training data. In order to reduce the impact of the training–testing discrepancy on SR performance, researchers propose to exploit the internal information of the LR input to learn image-specific SR model as shown in Fig. \ref{Fig.7}. 

The ``Zero-Shot'' SR (ZSSR) developed by Shocher \emph{et al.} \cite{shocher2018zero} is one of the representatives. \footnote{The source code of ZSSR \cite{shocher2018zero} is available at \url{https://github.com/assafshocher/ZSSR}} The self-supervised approach ZSSR \cite{shocher2018zero} is based on the cross-scale internal recurrence of information, which is a common property of natural images. More specifically, an eight-layer CNN is trained to model image-specific LR-HR relations in the testing phase, using the example pairs extracted from the LR test image and its degraded version. In consideration of the insufficiency of training data (the test image only), data augmentation is adopted when extracting image-specific LR-HR pairs. Since ZSSR \cite{shocher2018zero} can adapt itself to different testing images, it achieves excellent SR performance on real-world images, whose degradation process is non-ideal and unknown. Again based on the cross-scale recurrence property, Bell-Kligler \emph{et al.} \cite{bell2019blind} propose to train an image-specific GAN (KernelGAN) to model the degradation process (\emph{i.e.}, the blur kernel) of the input.\textsuperscript{\ref{FNKernelGAN}} Therefore, a fully self-supervised image-specific RSISR framework can be achieved when the blur kernel estimation module KernelGAN \cite{bell2019blind} is plugged into the reconstruction module ZSSR \cite{shocher2018zero}.  To jointly train the image-specific degradation and SR networks, Kim \emph{et al.} \cite{kim2020dual} design a unified internal learning-based SR framework DBPI, consisting of an SR network and a downscaling network. \footnote{The source code of DBPI \cite{kim2020dual} is available at \url{https://github.com/prote376/DBPI-BlindSR}} In the self-supervised training phase of DBPI, the SR network is optimized to reconstruct the LR input image from the its downscaled version produced by the downscaling network. Meanwhile, the downscaling network is trained to recover the LR input image from its super-resolved version generated by the SR network. Similarly, Emad \emph{et al.} \cite{emad2021dualsr} propose the DualSR that jointly optimizes an image-specific downsampler and corresponding upsampler. More specifically, DualSR \cite{emad2021dualsr} is trained with the cycle-consistency loss, the masked interpolation loss, and the adversarial loss using the patches from the test image. Results in \cite{kim2020dual, emad2021dualsr} show that the complementary training of the image-specific degradation and SR networks is beneficial to the reconstruction performance.

Although self-learning-based RSISR approaches such as ZSSR \cite{shocher2018zero}, KernelGAN \cite{bell2019blind}, and DBPI \cite{kim2020dual} can be easily adapted to LR input images, they generally have two main shortcomings due to the self-supervised training strategy. First, the optimization of SR models only utilizes the internal information of the LR input, while a great deal of external information is neglected. Second, these methods are usually time-consuming in the testing phase because of online training. To overcome these disadvantages, meta-learning is introduced into recent self-learning-based SR methods \cite{soh2020meta, park2020fast}. Based on ZSSR \cite{shocher2018zero}, Soh \emph{et al.} \cite{soh2020meta} present the meta-transfer learning for zero-shot SR (MZSR), which consists of three steps, \emph{i.e.}, large-scale training, meta-transfer learning, and meta-test. \footnote{The source code of MZSR \cite{soh2020meta} is available at \url{https://www.github.com/JWSoh/MZSR}} In order to ease the training of the SR network and the meta-learning, the large-scale training step first trains an eight-layer SR network with the pixel-wise $l_1$ loss on the large-scale dataset DIV2K \cite{agustsson2017ntire}. The meta-transfer learning process aims to find a generic initial point for internal learning following the Model-Agnostic Meta-Learning \cite{finn2017model}, making the model can be quickly adapted to new image conditions within a few gradient updates. In the meta-test phase, the input test image is first degraded to produce example pairs for model parameter update, and then it is fed into the updated model to generate an SR result. Thanks to the meta-transfer learning strategy, MZSR \cite{soh2020meta} achieves competitive performance in terms of both the quality of the super-resolved image and the running time. In \cite{park2020fast}, Park \emph{et al.} also propose to improve the performance of SOTA SR networks such as RCAN \cite{zhang2018image} using the meta-learning strategy, without changing the original architectures. \footnote{The source code of MLSR \cite{park2020fast} is available at \url{https://github.com/parkseobin/MLSR}}  On the whole, meta-learning-based SR approaches have strengths in reconstruction quality, generalization capability, and processing efficiency.

\begin{figure*}[!tb]
	\centering
	\subfigure[Ground-truth]{
		\includegraphics[width=4cm]{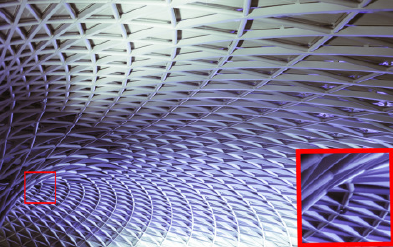}}
	\subfigure[Bicubic]{
		\includegraphics[width=4cm]{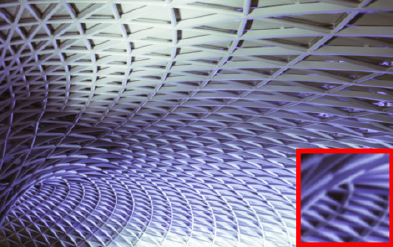}}
	\subfigure[ZSSR \cite{shocher2018zero}]{
		\includegraphics[width=4cm]{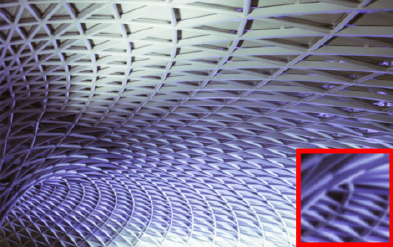}}
	\subfigure[ KernelGAN \cite{bell2019blind}]{
		\includegraphics[width=4cm]{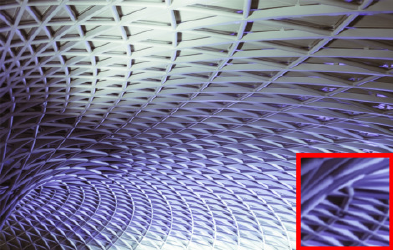}}\\
	\subfigure[MZSR (1) \cite{soh2020meta}]{
		\includegraphics[width=4cm]{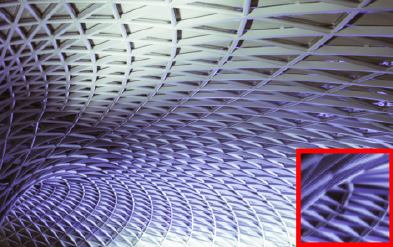}}
	\subfigure[DBPI \cite{kim2020dual}]{
		\includegraphics[width=4cm]{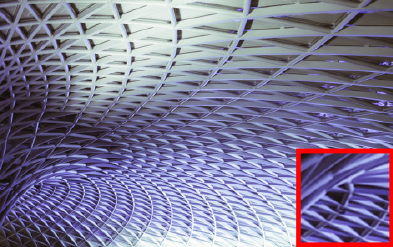}}
	\subfigure[ DAN  \cite{gu2019blind}]{
		\includegraphics[width=4cm]{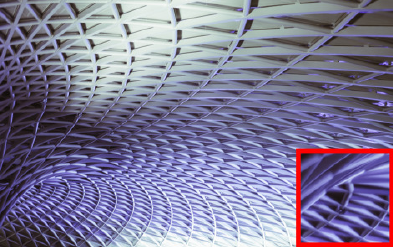}}
	\subfigure[SRResCGAN \cite{muhammad2020deep}]{
		\includegraphics[width=4cm]{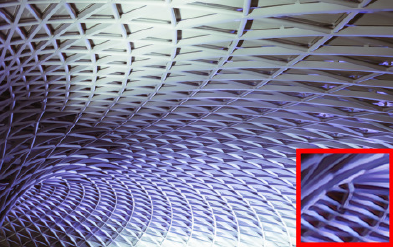}}

	\caption{Super-resolution results comparison ($\times 2$) on the image taken from DIV2KRK \cite{bell2019blind}.}
	\label{Fig.8}
\end{figure*}

\begin{figure*}[!tb]
	\centering
	\subfigure[Ground-truth]{
		\includegraphics[width=4cm]{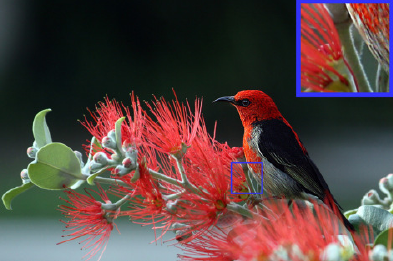}}
	\subfigure[Bicubic]{
		\includegraphics[width=4cm]{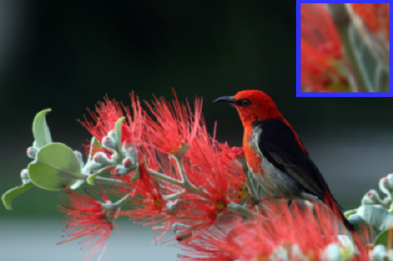}}
	\subfigure[ZSSR \cite{shocher2018zero}]{
		\includegraphics[width=4cm]{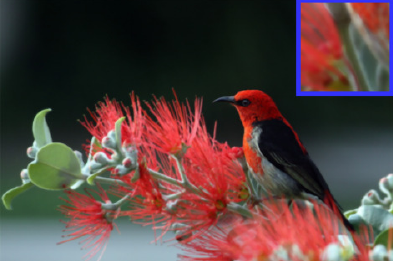}}
	\subfigure[ KernelGAN \cite{bell2019blind}]{
		\includegraphics[width=4cm]{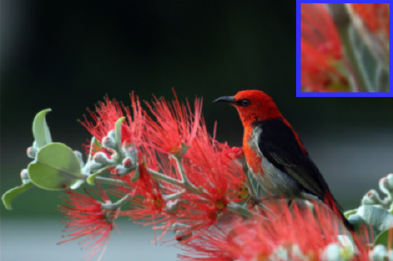}}\\
	\subfigure[DBPI \cite{kim2020dual}]{
		\includegraphics[width=4cm]{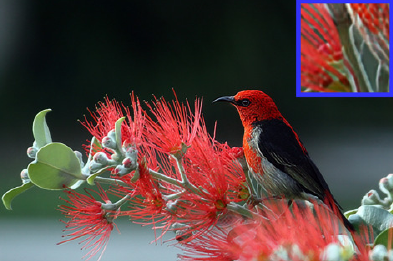}}
	\subfigure[DAN  \cite{huang2020unfolding}]{
		\includegraphics[width=4cm]{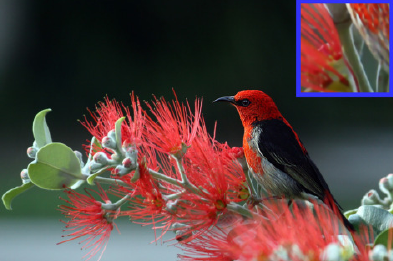}}
	\subfigure[IKC \cite{huang2020unfolding}]{
		\includegraphics[width=4cm]{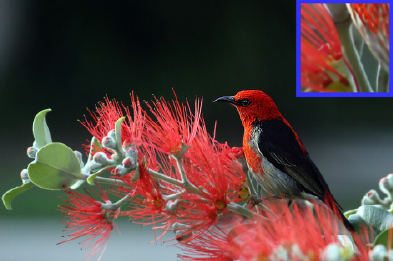}}
	\subfigure[SRResCGAN \cite{muhammad2020deep}]{
		\includegraphics[width=4cm]{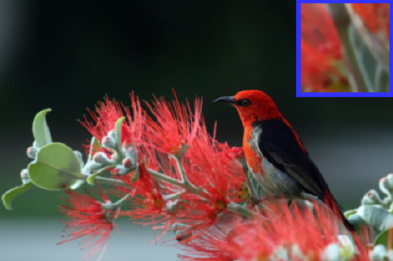}}
	
	\caption{Super-resolution results comparison ($\times 4$) on the image taken from DIV2KRK \cite{bell2019blind}.}
	\label{Fig.9}
\end{figure*}

\begin{figure*}[!tb]
	\centering
	\subfigure[Ground-truth]{
		\includegraphics[width=4cm]{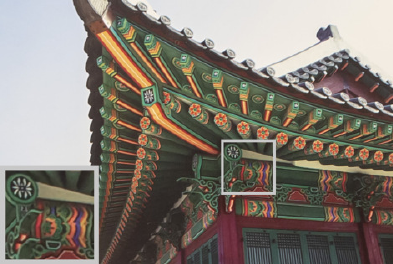}}
	\subfigure[Bicubic]{
		\includegraphics[width=4cm]{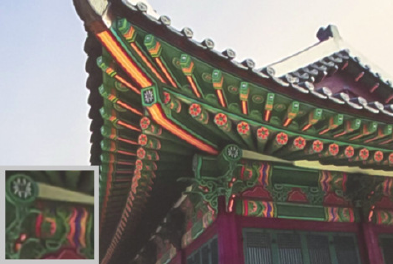}}
	\subfigure[ZSSR \cite{shocher2018zero}]{
		\includegraphics[width=4cm]{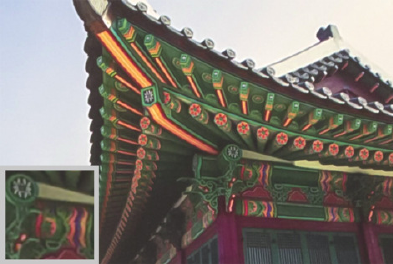}}
	\subfigure[ KernelGAN \cite{bell2019blind}]{
		\includegraphics[width=4cm]{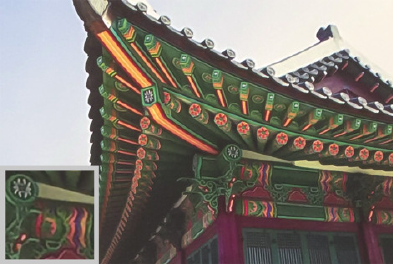}}\\
	\subfigure[MZSR  (1)  \cite{soh2020meta}]{
		\includegraphics[width=4cm]{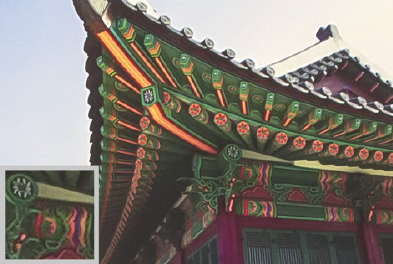}}
	\subfigure[DBPI \cite{kim2020dual}]{
		\includegraphics[width=4cm]{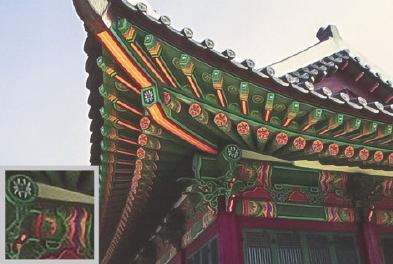}}
	\subfigure[ DAN  \cite{gu2019blind}]{
		\includegraphics[width=4cm]{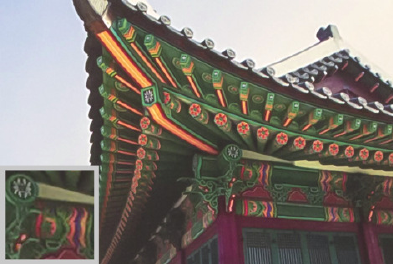}}
	\subfigure[SRResCGAN \cite{muhammad2020deep}]{
		\includegraphics[width=4cm]{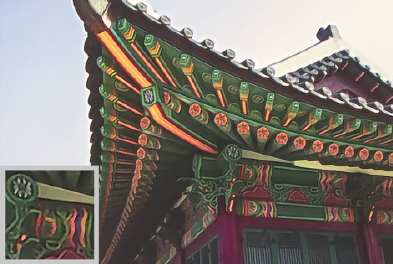}}
	
	\caption{Super-resolution results comparison ($\times 2$) on the image taken from RealSR \cite{cai2019toward}. }
	\label{Fig.10}
\end{figure*}

\begin{figure*}[!tb]
	\centering
	\subfigure[Ground-truth]{
		\includegraphics[width=4cm]{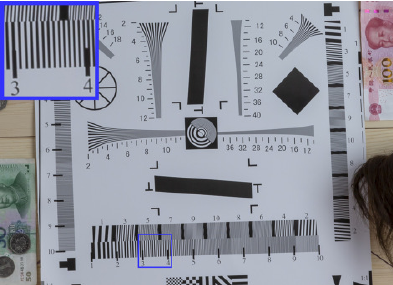}}
	\subfigure[Bicubic]{
		\includegraphics[width=4cm]{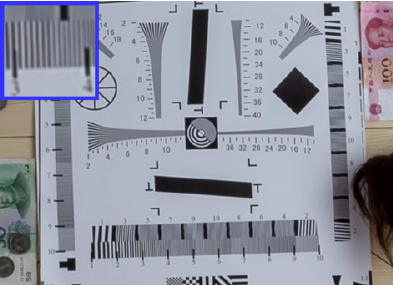}}
	\subfigure[ZSSR \cite{shocher2018zero}]{
		\includegraphics[width=4cm]{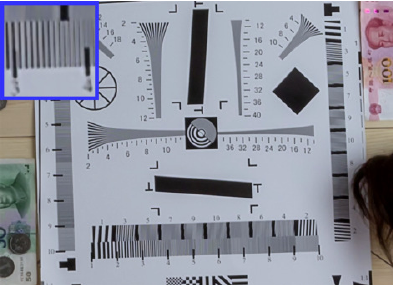}}
	\subfigure[ KernelGAN \cite{bell2019blind}]{
		\includegraphics[width=4cm]{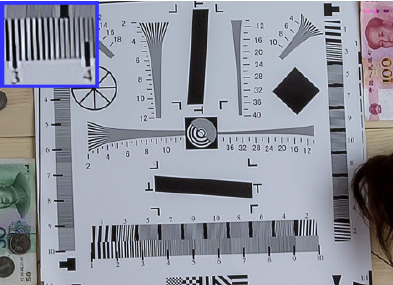}}\\
	\subfigure[DBPI \cite{kim2020dual}]{
		\includegraphics[width=4cm]{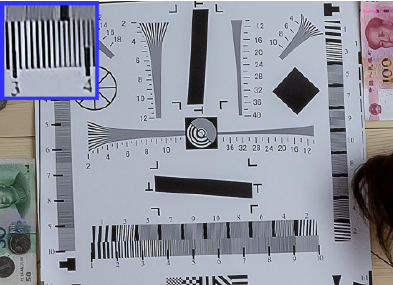}}
	\subfigure[DAN  \cite{huang2020unfolding}]{
		\includegraphics[width=4cm]{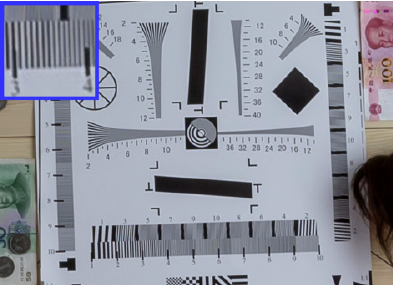}}
	\subfigure[IKC \cite{huang2020unfolding}]{
		\includegraphics[width=4cm]{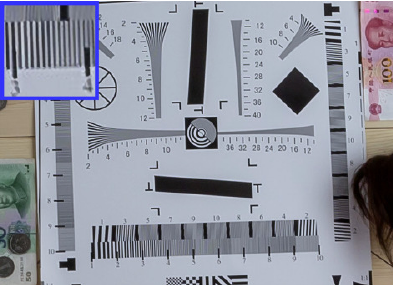}}
	\subfigure[SRResCGAN \cite{muhammad2020deep}]{
		\includegraphics[width=4cm]{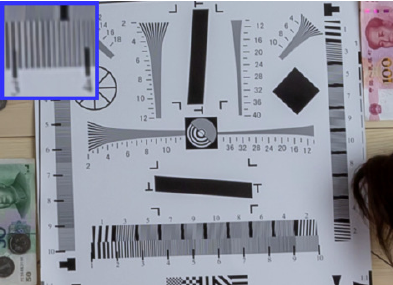}}
	
	\caption{Super-resolution results comparison ($\times 4$) on the image taken from RealSR \cite{cai2019toward}. }
	\label{Fig.11}
\end{figure*}

\begin{table*}[!t]
	\center
	\caption{The performance of representative RSISR algorithms on DIV2KRK \cite{bell2019blind} and RealSR \cite{cai2019toward} datasets.}
	\scalebox{1}[1]{
		\begin{tabular}{|c|c|c|c|c|c|c|c|c|}
			\hline
	
			\emph{Method} &  \emph{Scale}  & PSNR $\uparrow$  (dB)        & SSIM $\uparrow$ \cite{wang2004image}          & IFC $\uparrow$ \cite{sheikh2005information}           & NIQE $\downarrow$ \cite{mittal2012making}        & PIQE $\downarrow$ \cite{venkatanath2015blind}        & NRQM  $\uparrow$ \cite{ma2017learning}     & LPIPS $\downarrow$ \cite{zhang2018unreasonable}     \\ \hline
			
			\multicolumn{9}{|c|}{\emph{Performance on the synthetic dataset DIV2KRK \cite{bell2019blind}}} 
			\\  \hline
			
			\multirow{2}{*}{Bicubic}      & $\times 2$  &  27.24 &0.7846 &3.022 &5.196 &79.90 &3.273 &0.3631  \\ \cline{2-9} 
			& $\times 4$  &   23.89 &0.6478 &1.092 &6.310 &95.26 &3.043 &0.5645  \\ \hline
			
			\multirow{2}{*}{ZSSR  \cite{shocher2018zero}}      & $\times 2$ & 27.51 &0.7925 &3.000 &5.068 &80.06 &3.435 &0.3477 \\ \cline{2-9} 
			& $\times 4$  &  24.05 &0.6550 &1.132 &5.824 &90.29 &3.109 &0.5257  \\ \hline
			
			\multirow{2}{*}{KernelGAN  \cite{bell2019blind}} & $\times 2$ &         28.84 &0.8379 &3.720 &5.099 &72.17 &3.920 &0.2929         \\ \cline{2-9} 
			&  $\times 4$   &  24.76 &0.6799 &1.245 &5.886 &88.72 &3.610 &0.4980  \\ \hline
			
			MZSR (1) \cite{soh2020meta}                      & $\times 2$  &   26.69 &0.7889 &2.515 &5.627 &35.97 &4.112 &0.2630
			     \\ \hline
			
			\multirow{2}{*}{DBPI  \cite{kim2020dual}} & $\times 2$ &  29.55 &0.8657 &4.084 &5.239 &50.61 &4.844 &0.2641     \\ \cline{2-9} 
			&  $\times 4$   &   24.92 &0.7035 &1.385 &5.163 &70.74 &4.980 &0.4039  \\ \hline
			
			\multirow{2}{*}{DAN  \cite{huang2020unfolding}} & $\times 2$  &  31.06 &0.8848 &5.076 &4.041 &56.14 &3.519 &0.1667   \\ \cline{2-9} 
			&  $\times 4$   & 26.07 &0.7305 &1.758 &5.570 &82.28 &3.157 &0.4045 \\ \hline
			
			IKC  \cite{gu2019blind} & $\times 4$  & 25.41 &0.7255 &1.691 &5.140 &79.72 &3.869 &0.3977   \\ \cline{2-9} 
			\hline

			\multirow{2}{*}{SRResCGAN \cite{muhammad2020deep}}  & $\times 2$ &   26.15 &0.7486 &1.808 &3.717 &35.88 &4.976 &0.2463
           \\ \cline{2-9} 
			&  $\times 4$  &    24.00 &0.6497 &1.024 &5.038 &74.27 &3.054 &0.5054  \\ \hline

			\multicolumn{9}{|c|}{\emph{Performance on the real-world dataset  RealSR \cite{cai2019toward}}} 
			\\  \hline
			\multirow{2}{*}{Bicubic}      & $\times 2$  & 30.27 &0.8736 &1.921 &5.569 &83.90 &2.731 &0.2095 \\ \cline{2-9} 
			& $\times 4$  &   25.74 &0.7413 &0.890 &6.228 &92.53 &2.802 &0.4666  \\ \hline
			
			\multirow{2}{*}{ZSSR  \cite{shocher2018zero}}      & $\times 2$ &  30.56 &0.8786 &1.949 &5.376 &81.09 &2.749 &0.1756 \\ \cline{2-9} 
			& $\times 4$  &   25.83 &0.7434 &0.897 &4.971 &83.01 &2.797 &0.3503 \\ \hline
			
			\multirow{2}{*}{KernelGAN  \cite{bell2019blind}} & $\times 2$&  30.24 &0.8907 &2.106 &5.384 &77.10 &2.769 &0.1338       \\ \cline{2-9} 
			&  $\times 4$   &  24.09 &0.7243 &0.897 &4.918 &78.96 &3.559 &0.2981 \\ \hline
			
			MZSR (1) \cite{soh2020meta}                      & $\times 2$  &   27.96 &0.8160 &1.520 &5.887 &38.83 &2.868 &0.2105    \\ \hline
			
			\multirow{2}{*}{DBPI  \cite{kim2020dual}} & $\times 2$ &  27.86 &0.8285 &1.876 &5.698 &54.87 &2.995 &0.1777    \\ \cline{2-9} 
			&  $\times 4$   & 22.36 &0.6562 &0.851 &5.640 &70.58 &5.056 &0.3106
			   \\ \hline
			
			\multirow{2}{*}{DAN  \cite{huang2020unfolding}} & $\times 2$  &  30.63 &0.8815 &1.959 &4.387 &68.68 &2.745 &0.1314    \\ \cline{2-9} 
			&  $\times 4$   &   26.20  &0.7598 &0.966 &6.096 &90.57 &2.834 &0.4095
			
			  \\ \hline
			
			IKC  \cite{gu2019blind} & $\times 4$  &     25.60  &0.7488 &0.944 &4.845 &82.75 &2.927 &0.3188
			 \\ \cline{2-9} 

			\hline
			\multirow{2}{*}{SRResCGAN \cite{muhammad2020deep}}  & $\times 2$ & 26.26 &0.7983 &1.573 &3.786 &39.40 &4.129 &0.2090     \\ \cline{2-9} 
			&  $\times 4$  &  25.84 &0.7459 &0.900 &5.009 &74.45 &2.795 &0.3746  \\ \hline
			
	\end{tabular}}
	\label{Tab.4}
\end{table*}

\section{Comparisons among State-of-the-Arts}
\label{S.CSoTA}

\begin{figure}[!tb]
	\centering
	\includegraphics[width = 8 cm]{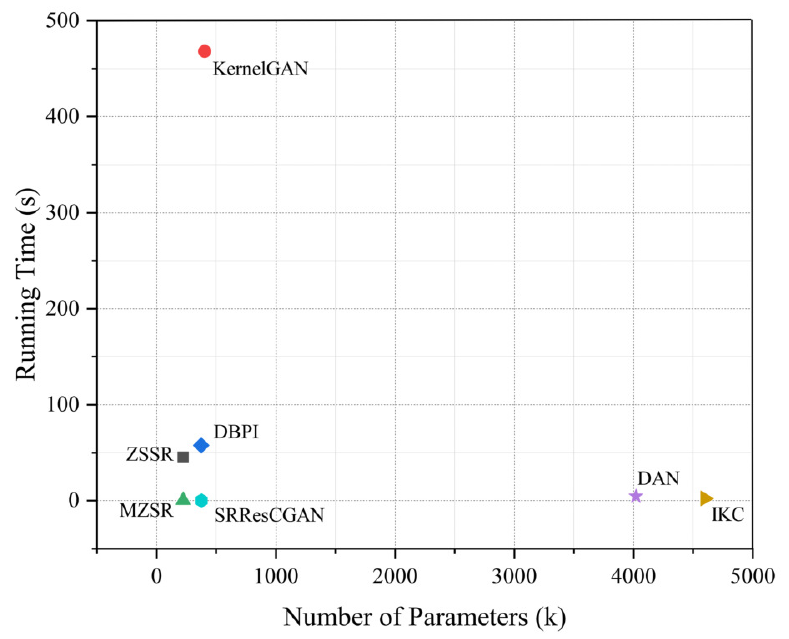}
	\caption{The number of parameters and the average running time on the images taken from RealSR \cite{cai2019toward}.}
	\label{Fig.12}
\end{figure}

In this section, we compare representative RSISR methods on benchmark datasets. More specifically, the selected competitors are ZSSR \cite{shocher2018zero}, KernelGAN \cite{bell2019blind}, MZSR \cite{soh2020meta}, DBPI \cite{kim2020dual}, DAN \cite{huang2020unfolding}, IKC \cite{gu2019blind}, and SRResCGAN \cite{muhammad2020deep}, covering multiple kinds of approaches mentioned in Section \ref{S.TechMe}. For these SR approaches, we use the official models provided by authors in the comparison, and the upsampling factor is set to $2$ or $4$. Two benchmark datasets are tested, including the DIV2KRK \cite{bell2019blind} and RealSR \cite{cai2019toward}. As introduced in Section \ref{S.DataSet}, DIV2KRK \cite{bell2019blind} is a synthetic dataset with random kernels and RealSR \cite{cai2019toward} is a realistic dataset collected by adjusting the focal length. \footnote{There are two subsets for testing in RealSR \cite{cai2019toward}, \emph{i.e.}, Canon and Nikon. We only use the 50 testing images in Canon in this comparison. } In addition to the subjective quality comparison presented in Figs. \ref{Fig.8}-\ref{Fig.11}, the SR results by different methods are objectively evaluated in Table \ref{Tab.4} on full-reference and no-reference image quality assessment metrics, including PSNR,  SSIM \cite{wang2004image},  IFC \cite{sheikh2005information},  LPIPS \cite{zhang2018unreasonable},  NIQE \cite{mittal2012making},  PIQE \cite{venkatanath2015blind},  and NRQM \cite{ma2017learning}. Beyond reconstruction accuracy, the model size and execution speed are also significant for SR algorithms. Therefore, the number of parameters and running time are also presented in Fig. \ref{Fig.12}. It is worth noting that the results shown in this section may be different from those in the original paper due to different settings of test environments, hyper-parameters, \emph{etc}. The aim of these comparisons is not to find a winner in terms of accuracy or efficiency, but to indicate the current state of the research on RSISR. In fact, as is known to all, it is not easy to make a completely fair comparison among these competitors due to the complexity of settings. As far as we know, in particular, there is no uniform or universally accepted settings yet for the comparison of RSISR models. 

We can make the following observations from the comparison results on visual quality, objective quality, and complexity. (\romannumeral1)~Compared with conventional interpolation (\emph{e.g.}, ``Bicubic''), SR is undoubtedly a more effective way to obtain HR images. (\romannumeral2) Overall, there is an obvious gap between the super-resolved image and the corresponding ground-truth in terms of visual effects, especially for texture and edge regions. For example, some SR results suffer from over-smoothing or over-sharpening artifacts. (\romannumeral3) There are some differences between subjective and objective assessment results. Moreover, the scores by different objective assessment metrics may not be necessarily consistent with each other. (\romannumeral4) In general, self-supervised learning-based SR approaches have fewer parameters than the SR models trained on large-scale datasets, but take longer to produce upsampled images. 

\section{Current Challenges and Future Directions}
\label{S.FuDir}

The numerous studies reviewed in Sections \ref{S.DataEvalMe} and \ref{S.TechMe} demonstrate the great progress of the research on RSISR. In fact, however, there are still problems requiring further exploration. We discuss some of the challenges and promising directions for RSISR in this section.

\subsection{Image Datasets}
\label{S.RWDataset}
Overall, the data for SR model training may be equally as crucial as the SR techniques for the study on RSISR, especially for deep learning-based solutions. Several realistic datasets have been constructed in the past few years, significantly boosting the reconstruction performance. However, compared with the datasets for popular computer vision tasks such as classification and detection, the lack of realistic datasets for RSISR is still striking. Therefore, it is desired to build larger and more representative/targeted realistic datasets for RSISR with considering the limiting factors of imaging resolution in the future. Meanwhile, more accurate alignment of the images captured from the same scene with different resolutions is also needed.

\subsection{SR Algorithms}
\label{S.Algori}
Although the SR performance on real-world images is getting better and better, there is still a long way to go before applying RSISR algorithms to practical applications. First, the LR images captured in real scenarios are likely to suffer from distinctly different degradations, challenging existing RSISR algorithms. Therefore, it is necessary to make RSISR models can adapt themselves to ever-changing real-world images. Second, most existing RSISR approaches are with a large model (\emph{e.g.}, a deep network), thereby requiring vast computing resources/time for forward inference and plenty of space for parameters storage. However, these resources are generally limited in real-world applications. Hence, how to achieve lightweight design and implementation of SR models without significant performance degradation is a primary challenge. Third, it is hard to obtain paired training data, even unpaired data or relevant HR references in some cases. Thus it is promising to develop RSISR models that can be run with unpaired training data or even the LR input only. Moreover,  how to leverage fewer models to meet personalized and multifunctional demands (\emph{e.g.}, the need for arbitrary upscaling factors and the preference to perceptual quality) of users deserves further investigation.

\subsection{Evaluation Criteria}
\label{S.Algori}

The evaluation criteria are vitally crucial for the research on computer vision tasks. On the one hand, the design of objective functions is generally guided by the evaluation criteria. For example, the $l_2$ loss is prevalent for image/video restoration tasks because it is highly correlated to the commonly used quality assessment metric PSNR. On the other hand, evaluation criteria are needed to make comparisons among different approaches, thus continuously advancing techniques. As previously mentioned, currently the PSNR and SSIM are the two of the most popular evaluation metrics for SR. However, previous studies show that they are unable to measure the visual quality of super-resolved images accurately. In addition, PSNR and SSIM are full-reference evaluation metrics that cannot be adopted in practical applications. Therefore, developing more suitable evaluation criteria for RSISR is a crucial and urgent research problem. On the whole, task-specific evaluation metrics are needed. That is, we have to take the goals and characteristics of RSISR into consideration while developing evaluation criteria. For example, the common targets include smoothness preserving for flat areas, detail enhancing for textures, sharpening for edges, \emph{etc}. Correspondingly, an evaluation criterion that covers these region-differentiated goals is desired. Meanwhile, the SR is performed generally for better human visual perception. Therefore, how to develop an automatic model that can measure visual quality more accurately and conveniently remains a challenge. Furthermore, there are no HR images for reference in practical applications. Thus the no-reference quality assessment criteria for super-resolved images have great demands.

\section{Conclusion}
\label{S.Conclu}
In recent years the super-resolution of real-world images has been getting increased attention. This paper briefly reviews recent super-resolution methods for realistic images, including degradation modeling-based algorithms, image pairs-based algorithms, domain translation-based algorithms, and self-learning-based algorithms. Meanwhile, we summarize the commonly used datasets and assessment metrics for RSISR models training and evaluation. Moreover, although some progress has been made on RSISR in the past few years, we point out that there are still challenges to be further addressed, \emph{e.g.}, realistic datasets for model training and testing, specific models for real-world image super-resolution and reconstruction performance evaluation. These unsolved problems also indicate the promising directions for future exploration.  We expect that this review can give a better understanding of existing studies for researchers, and also hope that it can attract more attention to advance the progress and application of real-world image super-resolution techniques.

\bibliography{Reference}

\begin{thebibliography}{100}
\providecommand{\url}[1]{#1}
\csname url@samestyle\endcsname
\providecommand{\newblock}{\relax}
\providecommand{\bibinfo}[2]{#2}
\providecommand{\BIBentrySTDinterwordspacing}{\spaceskip=0pt\relax}
\providecommand{\BIBentryALTinterwordstretchfactor}{4}
\providecommand{\BIBentryALTinterwordspacing}{\spaceskip=\fontdimen2\font plus
\BIBentryALTinterwordstretchfactor\fontdimen3\font minus
  \fontdimen4\font\relax}
\providecommand{\BIBforeignlanguage}[2]{{%
\expandafter\ifx\csname l@#1\endcsname\relax
\typeout{** WARNING: IEEEtran.bst: No hyphenation pattern has been}%
\typeout{** loaded for the language `#1'. Using the pattern for}%
\typeout{** the default language instead.}%
\else
\language=\csname l@#1\endcsname
\fi
#2}}
\providecommand{\BIBdecl}{\relax}
\BIBdecl

\bibitem{dai2016image}
D.~Dai, Y.~Wang, Y.~Chen, and L.~Van~Gool, ``Is image super-resolution helpful
  for other vision tasks?'' in \emph{IEEE Winter Conference on Applications of
  Computer Vision (WACV)}, 2016, pp. 1--9.

\bibitem{lei2019simultaneous}
S.~Lei, Z.~Shi, X.~Wu, B.~Pan, X.~Xu, and H.~Hao, ``Simultaneous
  super-resolution and segmentation for remote sensing images,'' in \emph{IEEE
  International Geoscience and Remote Sensing Symposium (IGARSS)}, 2019, pp.
  3121--3124.

\bibitem{guo2019super}
Z.~Guo, G.~Wu, X.~Song, W.~Yuan, Q.~Chen, H.~Zhang, X.~Shi, M.~Xu, Y.~Xu,
  R.~Shibasaki \emph{et~al.}, ``Super-resolution integrated building semantic
  segmentation for multi-source remote sensing imagery,'' \emph{IEEE Access},
  vol.~7, pp. 99\,381--99\,397, 2019.

\bibitem{wang2020dual}
L.~Wang, D.~Li, Y.~Zhu, L.~Tian, and Y.~Shan, ``Dual super-resolution learning
  for semantic segmentation,'' in \emph{Proceedings of the IEEE Conference on
  Computer Vision and Pattern Recognition (CVPR)}, 2020, pp. 3774--3783.

\bibitem{haris2018task}
M.~Haris, G.~Shakhnarovich, and N.~Ukita, ``Task-driven super resolution:
  Object detection in low-resolution images,'' \emph{arXiv preprint
  arXiv:1803.11316}, 2018.

\bibitem{pang2019jcs}
Y.~Pang, J.~Cao, J.~Wang, and J.~Han, ``{JCS-Net}: Joint classification and
  super-resolution network for small-scale pedestrian detection in surveillance
  images,'' \emph{IEEE Transactions on Information Forensics and Security},
  vol.~14, no.~12, pp. 3322--3331, 2019.

\bibitem{zhang2020kgsnet}
Y.~Zhang, Y.~Bai, M.~Ding, S.~Xu, and B.~Ghanem, ``{KGSnet}: key-point-guided
  super-resolution network for pedestrian detection in the wild,'' \emph{IEEE
  Transactions on Neural Networks and Learning Systems}, 2020.

\bibitem{wang2016studying}
Z.~Wang, S.~Chang, Y.~Yang, D.~Liu, and T.~S. Huang, ``Studying very low
  resolution recognition using deep networks,'' in \emph{Proceedings of the
  IEEE Conference on Computer Vision and Pattern Recognition (CVPR)}, 2016, pp.
  4792--4800.

\bibitem{yang2018long}
X.~Yang, W.~Wu, K.~Liu, P.~W. Kim, A.~K. Sangaiah, and G.~Jeon, ``Long-distance
  object recognition with image super resolution: A comparative study,''
  \emph{IEEE Access}, vol.~6, pp. 13\,429--13\,438, 2018.

\bibitem{suprapto2019gsr}
T.~Suprapto~Siadari, M.~Han, and H.~Yoon, ``{GSR-MAR}: Global super-resolution
  for person multi-attribute recognition,'' in \emph{Proceedings of the IEEE
  International Conference on Computer Vision Workshops (ICCVW)}, 2019.

\bibitem{belekos2010maximum}
S.~P. Belekos, N.~P. Galatsanos, and A.~K. Katsaggelos, ``Maximum a posteriori
  video super-resolution using a new multichannel image prior,'' \emph{IEEE
  Transactions on Image Processing}, vol.~19, no.~6, pp. 1451--1464, 2010.

\bibitem{liu2013bayesian}
C.~Liu and D.~Sun, ``On bayesian adaptive video super resolution,'' \emph{IEEE
  Transactions on Pattern Analysis and Machine Intelligence}, vol.~36, no.~2,
  pp. 346--360, 2013.

\bibitem{li2016video}
K.~Li, Y.~Zhu, J.~Yang, and J.~Jiang, ``Video super-resolution using an
  adaptive superpixel-guided auto-regressive model,'' \emph{Pattern
  Recognition}, vol.~51, pp. 59--71, 2016.

\bibitem{kappeler2016video}
A.~Kappeler, S.~Yoo, Q.~Dai, and A.~K. Katsaggelos, ``Video super-resolution
  with convolutional neural networks,'' \emph{IEEE Transactions on
  Computational Imaging}, vol.~2, no.~2, pp. 109--122, 2016.

\bibitem{caballero2017real}
J.~Caballero, C.~Ledig, A.~Aitken, A.~Acosta, J.~Totz, Z.~Wang, and W.~Shi,
  ``Real-time video super-resolution with spatio-temporal networks and motion
  compensation,'' in \emph{Proceedings of the IEEE Conference on Computer
  Vision and Pattern Recognition (CVPR)}, 2017, pp. 4778--4787.

\bibitem{jo2018deep}
Y.~Jo, S.~Wug~Oh, J.~Kang, and S.~Joo~Kim, ``Deep video super-resolution
  network using dynamic upsampling filters without explicit motion
  compensation,'' in \emph{Proceedings of the IEEE conference on Computer
  Vision and Pattern Recognition (CVPR)}, 2018, pp. 3224--3232.

\bibitem{lucas2019generative}
A.~Lucas, S.~Lopez-Tapia, R.~Molina, and A.~K. Katsaggelos, ``Generative
  adversarial networks and perceptual losses for video super-resolution,''
  \emph{IEEE Transactions on Image Processing}, vol.~28, no.~7, pp. 3312--3327,
  2019.

\bibitem{haris2019recurrent}
M.~Haris, G.~Shakhnarovich, and N.~Ukita, ``Recurrent back-projection network
  for video super-resolution,'' in \emph{Proceedings of the IEEE Conference on
  Computer Vision and Pattern Recognition (CVPR)}, 2019, pp. 3897--3906.

\bibitem{tian2020tdan}
Y.~Tian, Y.~Zhang, Y.~Fu, and C.~Xu, ``{TDAN}: Temporally-deformable alignment
  network for video super-resolution,'' in \emph{Proceedings of the IEEE
  Conference on Computer Vision and Pattern Recognition (CVPR)}, 2020, pp.
  3360--3369.

\bibitem{dikbas2012novel}
S.~Dikbas and Y.~Altunbasak, ``Novel true-motion estimation algorithm and its
  application to motion-compensated temporal frame interpolation,'' \emph{IEEE
  Transactions on Image Processing}, vol.~22, no.~8, pp. 2931--2945, 2012.

\bibitem{choi2007motion}
B.-D. Choi, J.-W. Han, C.-S. Kim, and S.-J. Ko, ``Motion-compensated frame
  interpolation using bilateral motion estimation and adaptive overlapped block
  motion compensation,'' \emph{IEEE Transactions on Circuits and Systems for
  Video Technology}, vol.~17, no.~4, pp. 407--416, 2007.

\bibitem{niklaus2017video1}
S.~Niklaus, L.~Mai, and F.~Liu, ``Video frame interpolation via adaptive
  separable convolution,'' in \emph{Proceedings of the IEEE International
  Conference on Computer Vision (ICCV)}, 2017, pp. 261--270.

\bibitem{niklaus2018context}
S.~Niklaus and F.~Liu, ``Context-aware synthesis for video frame
  interpolation,'' in \emph{Proceedings of the IEEE Conference on Computer
  Vision and Pattern Recognition (CVPR)}, 2018, pp. 1701--1710.

\bibitem{bao2019depth}
W.~Bao, W.-S. Lai, C.~Ma, X.~Zhang, Z.~Gao, and M.-H. Yang, ``Depth-aware video
  frame interpolation,'' in \emph{Proceedings of the IEEE Conference on
  Computer Vision and Pattern Recognition (CVPR)}, 2019, pp. 3703--3712.

\bibitem{peleg2019net}
T.~Peleg, P.~Szekely, D.~Sabo, and O.~Sendik, ``{IM-Net} for high resolution
  video frame interpolation,'' in \emph{Proceedings of the IEEE Conference on
  Computer Vision and Pattern Recognition (CVPR)}, 2019, pp. 2398--2407.

\bibitem{bao2019memc}
W.~Bao, W.-S. Lai, X.~Zhang, Z.~Gao, and M.-H. Yang, ``{MEMC-Net}: Motion
  estimation and motion compensation driven neural network for video
  interpolation and enhancement,'' \emph{IEEE Transactions on Pattern Analysis
  and Machine Intelligence}, 2019.

\bibitem{cheng2020video}
X.~Cheng and Z.~Chen, ``Video frame interpolation via deformable separable
  convolution.'' in \emph{AAAI}, 2020, pp. 10\,607--10\,614.

\bibitem{farsiu2004fast}
S.~Farsiu, M.~D. Robinson, M.~Elad, and P.~Milanfar, ``Fast and robust
  multiframe super resolution,'' \emph{IEEE Transactions on Image Processing},
  vol.~13, no.~10, pp. 1327--1344, 2004.

\bibitem{farsiu2005multiframe}
S.~Farsiu, M.~Elad, and P.~Milanfar, ``Multiframe demosaicing and
  super-resolution of color images,'' \emph{IEEE Transactions on Image
  Processing}, vol.~15, no.~1, pp. 141--159, 2005.

\bibitem{li2010multi}
X.~Li, Y.~Hu, X.~Gao, D.~Tao, and B.~Ning, ``A multi-frame image
  super-resolution method,'' \emph{Signal Processing}, vol.~90, no.~2, pp.
  405--414, 2010.

\bibitem{yuan2011multiframe}
Q.~Yuan, L.~Zhang, and H.~Shen, ``Multiframe super-resolution employing a
  spatially weighted total variation model,'' \emph{IEEE Transactions on
  Circuits and Systems for Video Technology}, vol.~22, no.~3, pp. 379--392,
  2011.

\bibitem{yue2014locally}
L.~Yue, H.~Shen, Q.~Yuan, and L.~Zhang, ``A locally adaptive l1- l2 norm for
  multi-frame super-resolution of images with mixed noise and outliers,''
  \emph{Signal Processing}, vol. 105, pp. 156--174, 2014.

\bibitem{kohler2016robust}
T.~K{\"o}hler, X.~Huang, F.~Schebesch, A.~Aichert, A.~Maier, and J.~Hornegger,
  ``Robust multiframe super-resolution employing iteratively re-weighted
  minimization,'' \emph{IEEE Transactions on Computational Imaging}, vol.~2,
  no.~1, pp. 42--58, 2016.

\bibitem{liu2018robust}
X.~Liu, L.~Chen, W.~Wang, and J.~Zhao, ``Robust multi-frame super-resolution
  based on spatially weighted half-quadratic estimation and adaptive btv
  regularization,'' \emph{IEEE Transactions on Image Processing}, vol.~27,
  no.~10, pp. 4971--4986, 2018.

\bibitem{laghrib2019new}
A.~Laghrib, A.~Hadri, A.~Hakim, and S.~Raghay, ``A new multiframe
  super-resolution based on nonlinear registration and a spatially weighted
  regularization,'' \emph{Information Sciences}, vol. 493, pp. 34--56, 2019.

\bibitem{sun2008image}
J.~Sun, Z.~Xu, and H.-Y. Shum, ``Image super-resolution using gradient profile
  prior,'' in \emph{IEEE Conference on Computer Vision and Pattern Recognition
  (CVPR)}, 2008, pp. 1--8.

\bibitem{zhang2012single}
K.~Zhang, X.~Gao, D.~Tao, and X.~Li, ``Single image super-resolution with
  non-local means and steering kernel regression,'' \emph{IEEE Transactions on
  Image Processing}, vol.~21, no.~11, pp. 4544--4556, 2012.

\bibitem{dong2012nonlocally}
W.~Dong, L.~Zhang, G.~Shi, and X.~Li, ``Nonlocally centralized sparse
  representation for image restoration,'' \emph{IEEE Transactions on Image
  Processing}, vol.~22, no.~4, pp. 1620--1630, 2013.

\bibitem{papyan2015multi}
V.~Papyan and M.~Elad, ``Multi-scale patch-based image restoration,''
  \emph{IEEE Transactions on Image Processing}, vol.~25, no.~1, pp. 249--261,
  2015.

\bibitem{jiang2016single}
J.~Jiang, X.~Ma, C.~Chen, T.~Lu, Z.~Wang, and J.~Ma, ``Single image
  super-resolution via locally regularized anchored neighborhood regression and
  nonlocal means,'' \emph{IEEE Transactions on Multimedia}, vol.~19, no.~1, pp.
  15--26, 2017.

\bibitem{ren2016single}
C.~Ren, X.~He, and T.~Q. Nguyen, ``Single image super-resolution via adaptive
  high-dimensional non-local total variation and adaptive geometric feature,''
  \emph{IEEE Transactions on Image Processing}, vol.~26, no.~1, pp. 90--106,
  2017.

\bibitem{chen2017single}
H.~Chen, X.~He, L.~Qing, and Q.~Teng, ``Single image super-resolution via
  adaptive transform-based nonlocal self-similarity modeling and learning-based
  gradient regularization,'' \emph{IEEE Transactions on Multimedia}, vol.~19,
  no.~8, pp. 1702--1717, 2017.

\bibitem{chang2019data}
K.~Chang, X.~Zhang, P.~L.~K. Ding, and B.~Li, ``Data-adaptive low-rank modeling
  and external gradient prior for single image super-resolution,'' \emph{Signal
  Processing}, vol. 161, pp. 36--49, 2019.

\bibitem{li2019single}
T.~Li, X.~Dong, and H.~Chen, ``Single image super-resolution incorporating
  example-based gradient profile estimation and weighted adaptive p-norm,''
  \emph{Neurocomputing}, vol. 355, pp. 105--120, 2019.

\bibitem{li2020adaptive}
J.~Li and W.~Guan, ``Adaptive $l_q$-norm constrained general nonlocal
  self-similarity regularizer based sparse representation for single image
  super-resolution,'' \emph{Information Fusion}, vol.~53, pp. 88--102, 2020.

\bibitem{freeman2002example}
W.~T. Freeman, T.~R. Jones, and E.~C. Pasztor, ``Example-based
  super-resolution,'' \emph{IEEE Computer Graphics and Applications}, vol.~22,
  no.~2, pp. 56--65, 2002.

\bibitem{glasner2009super}
D.~Glasner, S.~Bagon, and M.~Irani, ``Super-resolution from a single image,''
  in \emph{IEEE International Conference on Computer Vision (ICCV)}, 2009, pp.
  349--356.

\bibitem{xiong2013example}
Z.~Xiong, D.~Xu, X.~Sun, and F.~Wu, ``Example-based super-resolution with soft
  information and decision,'' \emph{IEEE Transactions on Multimedia}, vol.~15,
  no.~6, pp. 1458--1465, 2013.

\bibitem{zhu2014single}
Y.~Zhu, Y.~Zhang, and A.~L. Yuille, ``Single image super-resolution using
  deformable patches,'' in \emph{Proceedings of the IEEE Conference on Computer
  Vision and Pattern Recognition (CVPR)}, 2014, pp. 2917--2924.

\bibitem{huang2015single}
J.-B. Huang, A.~Singh, and N.~Ahuja, ``Single image super-resolution from
  transformed self-exemplars,'' in \emph{Proceedings of the IEEE Conference on
  Computer Vision and Pattern Recognition (CVPR)}, 2015, pp. 5197--5206.

\bibitem{li2016rotation}
T.~Li, X.~He, Q.~Teng, and X.~Wu, ``Rotation expanded dictionary-based single
  image super-resolution,'' \emph{Neurocomputing}, vol. 216, pp. 1--17, 2016.

\bibitem{huang2017srhrf+}
J.-J. Huang, T.~Liu, P.~Luigi~Dragotti, and T.~Stathaki, ``{SRHRF+}:
  Self-example enhanced single image super-resolution using hierarchical random
  forests,'' in \emph{Proceedings of the IEEE Conference on Computer Vision and
  Pattern Recognition Workshops (CVPRW)}, 2017, pp. 71--79.

\bibitem{yang2010image}
J.~Yang, J.~Wright, T.~S. Huang, and Y.~Ma, ``Image super-resolution via sparse
  representation,'' \emph{IEEE Transactions on Image Processing}, vol.~19,
  no.~11, pp. 2861--2873, 2010.

\bibitem{zeyde2010single}
R.~Zeyde, M.~Elad, and M.~Protter, ``On single image scale-up using
  sparse-representations,'' in \emph{International Conference on Curves and
  Surfaces}, 2010, pp. 711--730.

\bibitem{wang2012semi}
S.~Wang, L.~Zhang, Y.~Liang, and Q.~Pan, ``Semi-coupled dictionary learning
  with applications to image super-resolution and photo-sketch synthesis,'' in
  \emph{IEEE Conference on Computer Vision and Pattern Recognition (CVPR)},
  2012, pp. 2216--2223.

\bibitem{zhu2014fast}
Z.~Zhu, F.~Guo, H.~Yu, and C.~Chen, ``Fast single image super-resolution via
  self-example learning and sparse representation,'' \emph{IEEE Transactions on
  Multimedia}, vol.~16, no.~8, pp. 2178--2190, 2014.

\bibitem{kang2015learning}
L.-W. Kang, C.-C. Hsu, B.~Zhuang, C.-W. Lin, and C.-H. Yeh, ``Learning-based
  joint super-resolution and deblocking for a highly compressed image,''
  \emph{IEEE Transactions on Multimedia}, vol.~17, no.~7, pp. 921--934, 2015.

\bibitem{li2020combining}
X.~Li, G.~Cao, Y.~Zhang, A.~Shafique, and P.~Fu, ``Combining synthesis sparse
  with analysis sparse for single image super-resolution,'' \emph{Signal
  Processing: Image Communication}, vol.~83, p. 115805, 2020.

\bibitem{li2020depth}
B.~Li, Y.~Zhou, Y.~Zhang, and A.~Wang, ``Depth image super-resolution based on
  joint sparse coding,'' \emph{Pattern Recognition Letters}, vol. 130, pp.
  21--29, 2020.

\bibitem{ayas2020single}
S.~Ayas and M.~Ekinci, ``Single image super resolution using dictionary
  learning and sparse coding with multi-scale and multi-directional gabor
  feature representation,'' \emph{Information Sciences}, vol. 512, pp.
  1264--1278, 2020.

\bibitem{timofte2013anchored}
R.~Timofte, V.~De~Smet, and L.~Van~Gool, ``Anchored neighborhood regression for
  fast example-based super-resolution,'' in \emph{Proceedings of the IEEE
  International Conference on Computer Vision (ICCV)}, 2013, pp. 1920--1927.

\bibitem{timofte2014a+}
------, ``A+: Adjusted anchored neighborhood regression for fast
  super-resolution,'' in \emph{Asian Conference on Computer Vision (ACCV)},
  2014, pp. 111--126.

\bibitem{zhang2015joint}
K.~Zhang, B.~Wang, W.~Zuo, H.~Zhang, and L.~Zhang, ``Joint learning of multiple
  regressors for single image super-resolution,'' \emph{IEEE Signal Processing
  Letters}, vol.~23, no.~1, pp. 102--106, 2015.

\bibitem{agustsson2016regressor}
E.~Agustsson, R.~Timofte, and L.~Van~Gool, ``Regressor basis learning for
  anchored super-resolution,'' in \emph{2016 23rd International Conference on
  Pattern Recognition (ICPR)}, 2016, pp. 3850--3855.

\bibitem{perez2016antipodally}
E.~Perez-Pellitero, J.~Salvador, J.~Ruiz-Hidalgo, and B.~Rosenhahn,
  ``Antipodally invariant metrics for fast regression-based super-resolution,''
  \emph{IEEE Transactions on Image Processing}, vol.~25, no.~6, pp. 2456--2468,
  2016.

\bibitem{zhang2019learning}
K.~Zhang, Z.~Wang, J.~Li, X.~Gao, and Z.~Xiong, ``Learning recurrent residual
  regressors for single image super-resolution,'' \emph{Signal Processing},
  vol. 154, pp. 324--337, 2019.

\bibitem{dong2015image}
C.~Dong, C.~C. Loy, K.~He, and X.~Tang, ``Image super-resolution using deep
  convolutional networks,'' \emph{IEEE Transactions on Pattern Analysis and
  Machine Intelligence}, vol.~38, no.~2, pp. 295--307, 2015.

\bibitem{kim2016accurate}
J.~Kim, J.~Kwon~Lee, and K.~Mu~Lee, ``Accurate image super-resolution using
  very deep convolutional networks,'' in \emph{Proceedings of the IEEE
  Conference on Computer Vision and Pattern Recognition (CVPR)}, 2016, pp.
  1646--1654.

\bibitem{ledig2017photo}
C.~Ledig, L.~Theis, F.~Husz{\'a}r, J.~Caballero, A.~Cunningham, A.~Acosta,
  A.~Aitken, A.~Tejani, J.~Totz, Z.~Wang \emph{et~al.}, ``Photo-realistic
  single image super-resolution using a generative adversarial network,'' in
  \emph{Proceedings of the IEEE Conference on Computer Vision and Pattern
  Recognition (CVPR)}, 2017, pp. 4681--4690.

\bibitem{lim2017enhanced}
B.~Lim, S.~Son, H.~Kim, S.~Nah, and K.~Mu~Lee, ``Enhanced deep residual
  networks for single image super-resolution,'' in \emph{Proceedings of the
  IEEE Conference on Computer Vision and Pattern Recognition Workshops
  (CVPRW)}, 2017, pp. 136--144.

\bibitem{haris2018deep}
M.~Haris, G.~Shakhnarovich, and N.~Ukita, ``Deep back-projection networks for
  super-resolution,'' in \emph{Proceedings of the IEEE Conference on Computer
  Vision and Pattern Recognition (CVPR)}, 2018, pp. 1664--1673.

\bibitem{lai2018fast}
W.-S. Lai, J.-B. Huang, N.~Ahuja, and M.-H. Yang, ``Fast and accurate image
  super-resolution with deep laplacian pyramid networks,'' \emph{IEEE
  Transactions on Pattern Analysis and Machine Intelligence}, vol.~41, no.~11,
  pp. 2599--2613, 2018.

\bibitem{zhang2018image}
Y.~Zhang, K.~Li, K.~Li, L.~Wang, B.~Zhong, and Y.~Fu, ``Image super-resolution
  using very deep residual channel attention networks,'' in \emph{Proceedings
  of the European Conference on Computer Vision (ECCV)}, 2018, pp. 286--301.

\bibitem{zhang2018residual}
Y.~Zhang, Y.~Tian, Y.~Kong, B.~Zhong, and Y.~Fu, ``Residual dense network for
  image super-resolution,'' in \emph{Proceedings of the IEEE Conference on
  Computer Vision and Pattern Recognition (CVPR)}, 2018, pp. 2472--2481.

\bibitem{dai2019second}
T.~Dai, J.~Cai, Y.~Zhang, S.-T. Xia, and L.~Zhang, ``Second-order attention
  network for single image super-resolution,'' in \emph{Proceedings of the IEEE
  Conference on Computer Vision and Pattern Recognition (CVPR)}, 2019, pp.
  11\,065--11\,074.

\bibitem{guo2020closed}
Y.~Guo, J.~Chen, J.~Wang, Q.~Chen, J.~Cao, Z.~Deng, Y.~Xu, and M.~Tan,
  ``Closed-loop matters: Dual regression networks for single image
  super-resolution,'' in \emph{Proceedings of the IEEE Conference on Computer
  Vision and Pattern Recognition (CVPR)}, 2020, pp. 5407--5416.

\bibitem{2020Residual}
J.~Liu, W.~Zhang, Y.~Tang, J.~Tang, and G.~Wu, ``Residual feature aggregation
  network for image super-resolution,'' in \emph{Proceedings of the IEEE
  Conference on Computer Vision and Pattern Recognition (CVPR)}, 2020.

\bibitem{zhang2020gated}
X.~Zhang, H.~Dong, Z.~Hu, W.-S. Lai, F.~Wang, and M.-H. Yang, ``Gated fusion
  network for degraded image super resolution,'' \emph{International Journal of
  Computer Vision}, pp. 1--23, 2020.

\bibitem{kohler2019toward}
T.~K{\"o}hler, M.~B{\"a}tz, F.~Naderi, A.~Kaup, A.~Maier, and C.~Riess,
  ``Toward bridging the simulated-to-real gap: Benchmarking super-resolution on
  real data,'' \emph{IEEE Transactions on Pattern Analysis and Machine
  Intelligence}, vol.~42, no.~11, pp. 2944--2959, 2020.

\bibitem{cai2019toward}
J.~Cai, H.~Zeng, H.~Yong, Z.~Cao, and L.~Zhang, ``Toward real-world single
  image super-resolution: A new benchmark and a new model,'' in
  \emph{Proceedings of the IEEE International Conference on Computer Vision
  (ICCV)}, 2019, pp. 3086--3095.

\bibitem{wei2020component}
P.~Wei, Z.~Xie, H.~Lu, Z.~Zhan, Q.~Ye, W.~Zuo, and L.~Lin, ``Component
  divide-and-conquer for real-world image super-resolution,'' in \emph{European
  Conference on Computer Vision (ECCV)}, 2020.

\bibitem{chen2019camera}
C.~Chen, Z.~Xiong, X.~Tian, Z.-J. Zha, and F.~Wu, ``Camera lens
  super-resolution,'' in \emph{Proceedings of the IEEE Conference on Computer
  Vision and Pattern Recognition (CVPR)}, 2019, pp. 1652--1660.

\bibitem{zhang2019zoom}
X.~Zhang, Q.~Chen, R.~Ng, and V.~Koltun, ``Zoom to learn, learn to zoom,'' in
  \emph{Proceedings of the IEEE Conference on Computer Vision and Pattern
  Recognition (CVPR)}, 2019, pp. 3762--3770.

\bibitem{wang2020scene}
W.~Wang, E.~Xie, X.~Liu, W.~Wang, D.~Liang, C.~Shen, and X.~Bai, ``Scene text
  image super-resolution in the wild,'' in \emph{European Conference on
  Computer Vision (ECCV)}, 2020.

\bibitem{reza2020imagepairs}
H.~Reza Vaezi~Joze, I.~Zharkov, K.~Powell, C.~Ringler, L.~Liang, A.~Roulston,
  M.~Lutz, and V.~Pradeep, ``{ImagePairs}: Realistic super resolution dataset
  via beam splitter camera rig,'' in \emph{Proceedings of the IEEE Conference
  on Computer Vision and Pattern Recognition Workshops (CVPRW)}, 2020, pp.
  518--519.

\bibitem{xu2019towards}
X.~Xu, Y.~Ma, and W.~Sun, ``Towards real scene super-resolution with raw
  images,'' in \emph{Proceedings of the IEEE Conference on Computer Vision and
  Pattern Recognition (CVPR)}, 2019, pp. 1723--1731.

\bibitem{xu2020eploiting}
X.~Xu, Y.~Ma, W.~Sun, and M.-H. Yang, ``Exploiting raw images for real-scene
  super-resolution,'' \emph{IEEE Transactions on Pattern Analysis and Machine
  Intelligence}, 2020.

\bibitem{shao2015simple}
W.-Z. Shao and M.~Elad, ``Simple, accurate, and robust nonparametric blind
  super-resolution,'' in \emph{International Conference on Image and Graphics
  (ICIG)}, 2015, pp. 333--348.

\bibitem{shao2019nonparametric}
W.-Z. Shao, Q.~Ge, L.-Q. Wang, Y.-Z. Lin, H.-S. Deng, and H.-B. Li,
  ``Nonparametric blind super-resolution using adaptive heavy-tailed priors,''
  \emph{Journal of Mathematical Imaging and Vision}, vol.~61, no.~6, pp.
  885--917, 2019.

\bibitem{gu2019blind}
J.~Gu, H.~Lu, W.~Zuo, and C.~Dong, ``Blind super-resolution with iterative
  kernel correction,'' in \emph{Proceedings of the IEEE Conference on Computer
  Vision and Pattern Recognition (CVPR)}, 2019, pp. 1604--1613.

\bibitem{cornillere2019blind}
V.~Cornillere, A.~Djelouah, W.~Yifan, O.~Sorkine-Hornung, and C.~Schroers,
  ``Blind image super-resolution with spatially variant degradations,''
  \emph{ACM Transactions on Graphics}, vol.~38, no.~6, pp. 1--13, 2019.

\bibitem{huang2020unfolding}
Y.~Huang, S.~Li, L.~Wang, T.~Tan \emph{et~al.}, ``Unfolding the alternating
  optimization for blind super resolution,'' in \emph{34th Conference on Neural
  Information Processing Systems (NeurIPS)}, 2020.

\bibitem{michaeli2013nonparametric}
T.~Michaeli and M.~Irani, ``Nonparametric blind super-resolution,'' in
  \emph{Proceedings of the IEEE International Conference on Computer Vision
  (ICCV)}, 2013, pp. 945--952.

\bibitem{bell2019blind}
S.~Bell-Kligler, A.~Shocher, and M.~Irani, ``Blind super-resolution kernel
  estimation using an internal-gan,'' in \emph{33rd Conference on Neural
  Information Processing Systems (NeurIPS)}, 2019, pp. 284--293.

\bibitem{bulat2018learn}
A.~Bulat, J.~Yang, and G.~Tzimiropoulos, ``To learn image super-resolution, use
  a gan to learn how to do image degradation first,'' in \emph{Proceedings of
  the European Conference on Computer Vision (ECCV)}, 2018, pp. 185--200.

\bibitem{zhou2019kernel}
R.~Zhou and S.~Susstrunk, ``Kernel modeling super-resolution on real
  low-resolution images,'' in \emph{Proceedings of the IEEE International
  Conference on Computer Vision (ICCV)}, 2019, pp. 2433--2443.

\bibitem{xiao2020degradation}
J.~Xiao, H.~Yong, and L.~Zhang, ``Degradation model learning for real-world
  single image super-resolution,'' in \emph{Proceedings of the Asian Conference
  on Computer Vision (ACCV)}, 2020.

\bibitem{ji2020real}
X.~Ji, Y.~Cao, Y.~Tai, C.~Wang, J.~Li, and F.~Huang, ``Real-world
  super-resolution via kernel estimation and noise injection,'' in
  \emph{Proceedings of the IEEE Conference on Computer Vision and Pattern
  Recognition Workshops (CVPRW)}, 2020, pp. 466--467.

\bibitem{yuan2018unsupervised}
Y.~Yuan, S.~Liu, J.~Zhang, Y.~Zhang, C.~Dong, and L.~Lin, ``Unsupervised image
  super-resolution using cycle-in-cycle generative adversarial networks,'' in
  \emph{Proceedings of the IEEE Conference on Computer Vision and Pattern
  Recognition Workshops (CVPRW)}, 2018, pp. 701--710.

\bibitem{zhang2019multiple}
Y.~Zhang, S.~Liu, C.~Dong, X.~Zhang, and Y.~Yuan, ``Multiple cycle-in-cycle
  generative adversarial networks for unsupervised image super-resolution,''
  \emph{IEEE Transactions on Image Processing}, vol.~29, pp. 1101--1112, 2020.

\bibitem{kim2020unsupervised}
G.~Kim, J.~Park, K.~Lee, J.~Lee, J.~Min, B.~Lee, D.~K. Han, and H.~Ko,
  ``Unsupervised real-world super resolution with cycle generative adversarial
  network and domain discriminator,'' in \emph{Proceedings of the IEEE
  Conference on Computer Vision and Pattern Recognition Workshops (CVPRW)},
  2020, pp. 456--457.

\bibitem{maeda2020unpaired}
S.~Maeda, ``Unpaired image super-resolution using pseudo-supervision,'' in
  \emph{Proceedings of the IEEE Conference on Computer Vision and Pattern
  Recognition (CVPR)}, 2020, pp. 291--300.

\bibitem{prajapati2020unsupervised}
K.~Prajapati, V.~Chudasama, H.~Patel, K.~Upla, R.~Ramachandra, K.~Raja, and
  C.~Busch, ``Unsupervised single image super-resolution network ({USISResNet})
  for real-world data using generative adversarial network,'' in
  \emph{Proceedings of the IEEE Conference on Computer Vision and Pattern
  Recognition Workshops (CVPRW)}, 2020, pp. 464--465.

\bibitem{zhao2018unsupervised}
T.~Zhao, W.~Ren, C.~Zhang, D.~Ren, and Q.~Hu, ``Unsupervised degradation
  learning for single image super-resolution,'' \emph{arXiv preprint
  arXiv:1812.04240}, 2018.

\bibitem{you2019ct}
C.~You, G.~Li, Y.~Zhang, X.~Zhang, H.~Shan, M.~Li, S.~Ju, Z.~Zhao, Z.~Zhang,
  W.~Cong \emph{et~al.}, ``{CT} super-resolution gan constrained by the
  identical, residual, and cycle learning ensemble ({GAN-CIRCLE}),'' \emph{IEEE
  Transactions on Medical Imaging}, vol.~39, no.~1, pp. 188--203, 2020.

\bibitem{fritsche2019frequency}
M.~Fritsche, S.~Gu, and R.~Timofte, ``Frequency separation for real-world
  super-resolution,'' in \emph{IEEE International Conference on Computer Vision
  Workshop (ICCVW)}, 2019, pp. 3599--3608.

\bibitem{muhammad2020deep}
R.~Muhammad~Umer, G.~Luca~Foresti, and C.~Micheloni, ``Deep generative
  adversarial residual convolutional networks for real-world
  super-resolution,'' in \emph{Proceedings of the IEEE Conference on Computer
  Vision and Pattern Recognition Workshops (CVPRW)}, 2020, pp. 438--439.

\bibitem{rad2021benefiting}
M.~S. Rad, T.~Yu, C.~Musat, H.~K. Ekenel, B.~Bozorgtabar, and J.-P. Thiran,
  ``Benefiting from bicubically down-sampled images for learning real-world
  image super-resolution,'' in \emph{Proceedings of the IEEE Winter Conference
  on Applications of Computer Vision (WACV)}, 2021, pp. 1590--1599.

\bibitem{lugmayr2019unsupervised}
A.~Lugmayr, M.~Danelljan, and R.~Timofte, ``Unsupervised learning for
  real-world super-resolution,'' in \emph{IEEE International Conference on
  Computer Vision Workshop (ICCVW)}, 2019, pp. 3408--3416.

\bibitem{chen2020unsupervised}
S.~Chen, Z.~Han, E.~Dai, X.~Jia, Z.~Liu, L.~Xing, X.~Zou, C.~Xu, J.~Liu, and
  Q.~Tian, ``Unsupervised image super-resolution with an indirect supervised
  path,'' in \emph{Proceedings of the IEEE Conference on Computer Vision and
  Pattern Recognition Workshops (CVPRW)}, 2020, pp. 468--469.

\bibitem{shocher2018zero}
A.~Shocher, N.~Cohen, and M.~Irani, ``“{Zero-Shot}” super-resolution using
  deep internal learning,'' in \emph{Proceedings of the IEEE Conference on
  Computer Vision and Pattern Recognition (CVPR)}, 2018, pp. 3118--3126.

\bibitem{kim2020dual}
J.~Kim, C.~Jung, and C.~Kim, ``Dual back-projection-based internal learning for
  blind super-resolution,'' \emph{IEEE Signal Processing Letters}, vol.~27, pp.
  1190--1194, 2020.

\bibitem{emad2021dualsr}
M.~Emad, M.~Peemen, and H.~Corporaal, ``{DualSR}: Zero-shot dual learning for
  real-world super-resolution,'' in \emph{Proceedings of the IEEE Winter
  Conference on Applications of Computer Vision (WACV)}, 2021, pp. 1630--1639.

\bibitem{soh2020meta}
J.~W. Soh, S.~Cho, and N.~I. Cho, ``Meta-transfer learning for zero-shot
  super-resolution,'' in \emph{Proceedings of the IEEE Conference on Computer
  Vision and Pattern Recognition (CVPR)}, 2020, pp. 3516--3525.

\bibitem{park2020fast}
S.~Park, J.~Yoo, D.~Cho, J.~Kim, and T.~H. Kim, ``Fast adaptation to
  super-resolution networks via meta-learning,'' in \emph{European Conference
  on Computer Vision (ECCV)}, 2020.

\bibitem{ma2017learning}
C.~Ma, C.-Y. Yang, X.~Yang, and M.-H. Yang, ``Learning a no-reference quality
  metric for single-image super-resolution,'' \emph{Computer Vision and Image
  Understanding}, vol. 158, pp. 1--16, 2017.

\bibitem{fang2018blind}
Y.~Fang, C.~Zhang, W.~Yang, J.~Liu, and Z.~Guo, ``Blind visual quality
  assessment for image super-resolution by convolutional neural network,''
  \emph{Multimedia Tools and Applications}, vol.~77, no.~22, pp.
  29\,829--29\,846, 2018.

\bibitem{bare2018deep}
B.~Bare, K.~Li, B.~Yan, B.~Feng, and C.~Yao, ``A deep learning based
  no-reference image quality assessment model for single-image
  super-resolution,'' in \emph{2018 IEEE International Conference on Acoustics,
  Speech and Signal Processing (ICASSP)}.\hskip 1em plus 0.5em minus
  0.4em\relax IEEE, 2018, pp. 1223--1227.

\bibitem{greeshma2020super}
M.~Greeshma and V.~Bindu, ``Super-resolution quality criterion (srqc): a
  super-resolution image quality assessment metric,'' \emph{Multimedia Tools
  and Applications}, pp. 1--22, 2020.

\bibitem{cai2019ntire}
J.~Cai, S.~Gu, R.~Timofte, and L.~Zhang, ``{NTIRE} 2019 challenge on real image
  super-resolution: Methods and results,'' in \emph{Proceedings of the IEEE
  Conference on Computer Vision and Pattern Recognition Workshops (CVPRW)},
  2019, pp. 2211--2223.

\bibitem{lugmayr2019aim}
A.~Lugmayr, M.~Danelljan, R.~Timofte, M.~Fritsche, S.~Gu, K.~Purohit,
  P.~Kandula, M.~Suin, A.~Rajagoapalan, N.~H. Joon \emph{et~al.}, ``{AIM} 2019
  challenge on real-world image super-resolution: Methods and results,'' in
  \emph{IEEE International Conference on Computer Vision Workshop (ICCVW)},
  2019, pp. 3575--3583.

\bibitem{lugmayr2020ntire}
A.~Lugmayr, M.~Danelljan, and R.~Timofte, ``{NTIRE} 2020 challenge on
  real-world image super-resolution: Methods and results,'' in
  \emph{Proceedings of the IEEE Conference on Computer Vision and Pattern
  Recognition Workshops (CVPRW)}, 2020, pp. 494--495.

\bibitem{wei2020aim}
P.~Wei, H.~Lu, R.~Timofte, L.~Lin, W.~Zuo, Z.~Pan, B.~Li, T.~Xi, Y.~Fan,
  G.~Zhang \emph{et~al.}, ``{AIM} 2020 challenge on real image
  super-resolution: Methods and results,'' in \emph{European Conference on
  Computer Vision Workshops (ECCVW)}, 2020.

\bibitem{yue2016image}
L.~Yue, H.~Shen, J.~Li, Q.~Yuan, H.~Zhang, and L.~Zhang, ``Image
  super-resolution: The techniques, applications, and future,'' \emph{Signal
  Processing}, vol. 128, pp. 389--408, 2016.

\bibitem{yang2019deep}
W.~Yang, X.~Zhang, Y.~Tian, W.~Wang, J.-H. Xue, and Q.~Liao, ``Deep learning
  for single image super-resolution: A brief review,'' \emph{IEEE Transactions
  on Multimedia}, vol.~21, no.~12, pp. 3106--3121, 2019.

\bibitem{wang2020deep}
Z.~Wang, J.~Chen, and S.~C. Hoi, ``Deep learning for image super-resolution: A
  survey,'' \emph{IEEE Transactions on Pattern Analysis and Machine
  Intelligence}, 2020.

\bibitem{liu2020video}
H.~Liu, Z.~Ruan, P.~Zhao, F.~Shang, L.~Yang, and Y.~Liu, ``Video super
  resolution based on deep learning: A comprehensive survey,'' \emph{arXiv
  preprint arXiv:2007.12928}, 2020.

\bibitem{agustsson2017ntire}
E.~Agustsson and R.~Timofte, ``{NTIRE} 2017 challenge on single image
  super-resolution: Dataset and study,'' in \emph{Proceedings of the IEEE
  Conference on Computer Vision and Pattern Recognition Workshops (CVPRW)},
  2017, pp. 126--135.

\bibitem{arbelaez2011contour}
P.~Arbelaez, M.~Maire, C.~Fowlkes, and J.~Malik, ``Contour detection and
  hierarchical image segmentation,'' \emph{IEEE Transactions on Pattern
  Analysis and Machine Intelligence}, vol.~33, no.~5, pp. 898--916, 2011.

\bibitem{bevilacqua2012low}
M.~Bevilacqua, A.~Roumy, C.~Guillemot, and M.-L.~A. Morel, ``Low-complexity
  single-image super-resolution based on nonnegative neighbor embedding,'' in
  \emph{British Machine Vision Conference (BMVC)}, 2012.

\bibitem{fujimoto2016manga109}
A.~Fujimoto, T.~Ogawa, K.~Yamamoto, Y.~Matsui, T.~Yamasaki, and K.~Aizawa,
  ``Manga109 dataset and creation of metadata,'' in \emph{Proceedings of the
  1st International Workshop on Comics Analysis, Processing and Understanding},
  2016, pp. 1--5.

\bibitem{lowe2004distinctive}
D.~G. Lowe, ``Distinctive image features from scale-invariant keypoints,''
  \emph{International Journal of Computer Vision}, vol.~60, no.~2, pp. 91--110,
  2004.

\bibitem{fischler1981random}
M.~A. Fischler and R.~C. Bolles, ``Random sample consensus: a paradigm for
  model fitting with applications to image analysis and automated
  cartography,'' \emph{Communications of the ACM}, vol.~24, no.~6, pp.
  381--395, 1981.

\bibitem{evangelidis2008parametric}
G.~D. Evangelidis and E.~Z. Psarakis, ``Parametric image alignment using
  enhanced correlation coefficient maximization,'' \emph{IEEE Transactions on
  Pattern Analysis and Machine Intelligence}, vol.~30, no.~10, pp. 1858--1865,
  2008.

\bibitem{wang2004image}
Z.~Wang, A.~C. Bovik, H.~R. Sheikh, and E.~P. Simoncelli, ``Image quality
  assessment: from error visibility to structural similarity,'' \emph{IEEE
  Transactions on Image Processing}, vol.~13, no.~4, pp. 600--612, 2004.

\bibitem{sheikh2005information}
H.~R. Sheikh, A.~C. Bovik, and G.~De~Veciana, ``An information fidelity
  criterion for image quality assessment using natural scene statistics,''
  \emph{IEEE Transactions on Image Processing}, vol.~14, no.~12, pp.
  2117--2128, 2005.

\bibitem{zhang2018unreasonable}
R.~Zhang, P.~Isola, A.~A. Efros, E.~Shechtman, and O.~Wang, ``The unreasonable
  effectiveness of deep features as a perceptual metric,'' in \emph{Proceedings
  of the IEEE Conference on Computer Vision and Pattern Recognition (CVPR)},
  2018, pp. 586--595.

\bibitem{mittal2012making}
A.~Mittal, R.~Soundararajan, and A.~C. Bovik, ``Making a `completely blind'
  image quality analyzer,'' \emph{IEEE Signal Processing Letters}, vol.~20,
  no.~3, pp. 209--212, 2012.

\bibitem{venkatanath2015blind}
N.~Venkatanath, D.~Praneeth, M.~C. Bh, S.~S. Channappayya, and S.~S. Medasani,
  ``Blind image quality evaluation using perception based features,'' in
  \emph{2015 Twenty First National Conference on Communications (NCC)}, 2015,
  pp. 1--6.

\bibitem{yang2014single}
C.-Y. Yang, C.~Ma, and M.-H. Yang, ``Single-image super-resolution: A
  benchmark,'' in \emph{European Conference on Computer Vision (ECCV)}, 2014,
  pp. 372--386.

\bibitem{efrat2013accurate}
N.~Efrat, D.~Glasner, A.~Apartsin, B.~Nadler, and A.~Levin, ``Accurate blur
  models vs. image priors in single image super-resolution,'' in
  \emph{Proceedings of the IEEE International Conference on Computer Vision
  (ICCV)}, 2013, pp. 2832--2839.

\bibitem{mechrez2018contextual}
R.~Mechrez, I.~Talmi, and L.~Zelnik-Manor, ``The contextual loss for image
  transformation with non-aligned data,'' in \emph{Proceedings of the European
  Conference on Computer Vision (ECCV)}, 2018, pp. 768--783.

\bibitem{tomasi1998bilateral}
C.~Tomasi and R.~Manduchi, ``Bilateral filtering for gray and color images,''
  in \emph{IEEE International Conference on Computer Vision (ICCV)}, 1998, pp.
  839--846.

\bibitem{simonyan2015very}
K.~Simonyan and A.~Zisserman, ``Very deep convolutional networks for
  large-scale image recognition,'' in \emph{International Conference on
  Learning Representations (ICLR)}, 2015.

\bibitem{zhu2017unpaired}
J.-Y. Zhu, T.~Park, P.~Isola, and A.~A. Efros, ``Unpaired image-to-image
  translation using cycle-consistent adversarial networks,'' in
  \emph{Proceedings of the IEEE International Conference on Computer Vision
  (ICCV)}, 2017, pp. 2223--2232.

\bibitem{wang2018esrgan}
X.~Wang, K.~Yu, S.~Wu, J.~Gu, Y.~Liu, C.~Dong, Y.~Qiao, and C.~Change~Loy,
  ``{ESRGAN}: Enhanced super-resolution generative adversarial networks,'' in
  \emph{Proceedings of the European Conference on Computer Vision Workshops
  (ECCVW)}, 2018.

\bibitem{finn2017model}
C.~Finn, P.~Abbeel, and S.~Levine, ``Model-agnostic meta-learning for fast
  adaptation of deep networks,'' in \emph{International Conference on Machine
  Learning (ICML)}, 2017.

\end{thebibliography}
\bibliographystyle{IEEEtran}

\end{document}